\documentclass{ieeeaccess}

\usepackage{cite}
\usepackage{amsmath,amssymb,amsfonts}
\usepackage{algorithmic}
\usepackage{graphicx}
\usepackage{textcomp}
\def\BibTeX{{\rm B\kern-.05em{\sc i\kern-.025em b}\kern-.08em
    T\kern-.1667em\lower.7ex\hbox{E}\kern-.125emX}}

\usepackage{pgfplots}
\usepackage{pgfplotstable}
\pgfplotsset{compat=1.7}
\usepackage{tikz}

\NewSpotColorSpace{PANTONE}
\AddSpotColor{PANTONE} {PANTONE3015C} {PANTONE\SpotSpace 3015\SpotSpace C} {1 0.3 0 0.2}
\SetPageColorSpace{PANTONE}

\definecolor{IADB}{HTML}{0066A1}
\definecolor{IALB}{HTML}{019AD3}

\definecolor{burntorange}{cmyk}{19,15,16,0}
\definecolor{gry}{cmyk}{239,239,239,0}
\definecolor{greengashang}{HTML}{03AC13}
\tikzstyle{void}=[]
\usetikzlibrary{shapes,snakes,arrows, decorations.markings, mindmap, shadows}
\tikzstyle{newsuperpeers}=[draw,circle,burntorange, left color=blue!20, text=violet,minimum width=20pt]
\tikzstyle{legend_general}=[rectangle, rounded corners, thin, white, fill= white, draw, text=black, minimum width=2.5cm, minimum height=0.8cm]
\tikzstyle{superpeersgry}=[draw,circle,gry, left color=gry, text=violet,minimum width=20pt]
\tikzstyle{superpeersvio}=[draw,circle,black, left color=black!50, text=violet,minimum width=20pt]
\tikzstyle{newsuperpeersvio}=[draw,circle,black, left color=green!20, text=violet,minimum width=20pt]
\tikzstyle{superpeers}=[draw,circle,black, left color=black!20, text=violet,minimum width=20pt]
\tikzstyle{peers}=[draw,violet,bottom color=gry!30, top color= white, text=violet,minimum width=20pt]
\tikzstyle{Giantpeers}=[draw,circle,black, left color=black!20, text=violet,minimum width=100pt]

\begin{document}
		\bstctlcite{IEEEexample:BSTcontrol}
	
\history{Date of publication xxxx 00, 0000, date of current version xxxx 00, 0000.}
\doi{10.1109/ACCESS.2017.DOI}

\title{Handling State Space Explosion in Component-based Software Verification: A Review}
\author{\uppercase{Faranak Nejati}\authorrefmark{1}, 
\uppercase{Abdul Azim Abd. Ghani\authorrefmark{1}, Ng Keng Yap\authorrefmark{1}, and Azmi Jaafar}\authorrefmark{1}.}
\address[1]{FSKTM, Department of Software Engineering and Information systems, Universiti Putra Malaysia, 43400 UPM Serdang, Selangor Darul Ehsan, Malaysia (e-mail: faran.nejati@gmail.com, azim@upm.edu.my, kengyap@upm.edu.my, azmij@upm.edu.my)}


\corresp{Corresponding author: Abdul Azim Abd. Ghani (e-mail: azim@upm.edu.my)}

\begin{abstract}
Component-based software development (CBSD) is  an alternative approach to constructing software systems that offers numerous benefits, particularly in decreasing the complexity of system design. However, deploying components into a system is a challenging and error-prone task. Model-checking is one of the reliable methods to systematically analyze the correctness of a system. Its brute-force checking of the system's state space assists to significantly expand the level of confidence in the system. Nevertheless, model-checking is limited by a critical problem called state space explosion (SSE). To benefit from model-checking, an appropriate method is required to reduce SSE. In the past two decades, a great number of SSE reduction methods have been proposed containing many similarities, dissimilarities, and unclear concepts in some cases. This research, firstly, plans to present a review of SSE handling methods and classify them based on their similarities, principle, and characteristics. Second, it investigates the methods for handling SSE problem in the verification process of CBSD and provides insight into the potential limitations, underlining the key challenges for future research efforts.
\end{abstract}


\titlepgskip=-15pt

\maketitle

\section{Introduction}
 Component-based software development (CBSD) is a vital emerging topic in software engineering \cite{lau2007software, bachmann2000volume}. CBSD is an alternative approach of constructing systems from prebuilt software units (components) which offers numerous benefits, particularly in decreasing the complexity of the system design. However, deploying components into a system is a challenging and error-prone task. Errors may lead to destructive results. A single error can lead to an overall system crash, as in the error that crashed the Arian-5 rocket. There was a small application in Arian-5 for the inertial reference system that was trying to assign a 64-bit floating-point number into a variable with 16-bit space \cite{ehrlich1991faults}. This small mistake led to the catastrophic explosion of the aircraft. There are  various other safety-critical systems similar to the Arian-5 that could have disastrous outcomes if such errors occur; like in nuclear power stations, avionic software, aircraft flight control, and traffic control \cite{ogheneovo2014software}.

 Model-checking is one of the renowned approaches for verifying component-based software systems \cite{clarke1999model}. It is a brute-force verification method that is able to automatically and systematically analyze the specification and state space (SS) of a given system to demonstrate if its properties are satisfied completely or otherwise. This approach has been proposed independently by E. M. Clarke \textit{et al.} \cite{clarke1999model} and J. Sifakis \textit{et al.} \cite{queille1982specification}. The brute-force check of SS in model-checking significantly expands the level of confidence in the system.

However, model-checking is limited by state space explosion (SSE).  SSE occurs when a system's state space increases exponentially with the number of its components, thus rapidly surpasses the memory capacity of the computer. Subsequently, the amount of SS that can be checked by a model checker will be restricted. However, the promising advantages offered by model-checking have nevertheless encouraged the research community to tackle the SSE obstacle and  has spearheaded the major direction of model-checking research \cite{jhala2009software, tian2012abstract}. As a result, a massive collection of methods to alleviate SSE problems in all  domains of software development has been presented.  However, it should be determined which of these methods and algorithms would be sufficient in CBSD for supporting component-wise software development and its justification in order to ensure its compatibility in component-wise software systems.

In that effort, we first summarized in subsequent sections all the proposed  SSE reduction methods present in literature and provided a classification based on their principles and characteristics. The classification offers explanations pertaining to the key features and challenges in the literature that would aid in understanding model-checking and SSE reduction methods. The classification can then be utilized in the development of a new method that is suitable for a particular application domain.

 Additionally, it can be deduced further to determine from the classification all component-wise software methods that are utilized the most in CBSD to rectify its suitability for CBSD along with a discussion pertaining to the key feature and challenges that are mentioned in the literature.

To this end, we set up the following objectives: 1- reviewing and briefly describing the methods presented in the literature to address SSE problem in model checking; 2- classifying them based on their principles and characteristics; 3- deducing the key features and challenges of the methods  mentioned in the literature; 4- identifying and discussing the methods for tackling SSE problem used in CBSD in order to both analyze gaps and identify suitable methods for SSE reduction in CBSD. 

To complete these steps sufficiently, six research questions (RQs) have been formulated which are defined in Section~\ref{Section2}. Answering these questions aids in enhancing comprehension, determining the suitability of SSE reduction methods, and underlying the key features and challenges for future researches.

The presented study shares similarities with surveys presented in \cite{pelanek2008fighting, rafe2013survey, ACM6} for dealing with SSE problem, however, this paper collects the wide range of SSE reduction methods, offers explanations for each method, their success factor, and identifies challenges. Additionally, a discussion of SSE problems in component-based systems is also covered. Other review papers alike \cite{gabmeyer2013classification}, are based on the tool-sets which is beyond the scope of this paper.

The remainder of the paper is arranged as the following: in Section~\ref{Section2}, the research methodology is presented. Section~\ref{Section3} provides an explanation of the basic concepts about SSE used in this paper. Section~\ref{Section4} defines, classifies, and explains the different methods for alleviating SSE problem. In addition, it also contains a tabled summary of the key factors and  limitations of these methods. Section~\ref{result} contains a discussion about tackling SSE problem in CBSD and identifies the key challenges for future research. Section~\ref{Section6} discusses and concludes the results.

\section{Research method} \label{Section2}
In this section, the conducted research method is described. In order to collect studies; we carried out the following steps: \textit{(i)} formulating research questions, \textit{(ii)} identifying keywords, databases and search strategy, \textit{(iii)} obtaining inclusion and exclusion criteria to collect and analyse literatures, \textit{(iv)} quality assessment and selected studies, and \textit{(v)} information extraction.

\subsection{Research questions}
The objective of this work is to classify and briefly describe SSE reduction methods and the underlying key features and challenges of the methods in both general and CBSD systems. The set of corresponding research questions are listed in Table \ref{tab:RQ}. Answering these RQs establishes an effective overview of the most current SSE reduction method and  fulfills the objectives of this research.

\renewcommand{\arraystretch}{1.5}
\begin{table*}[h!]
	\caption{Research questions}
	\label{tab:RQ}
	\centering	
	\small
	\begin{tabular}{c p{7 cm} p{7 cm}}
		\hline
		\hline	 
		\textbf{NO} &     
		\textbf{Research questions} &      
		\textbf{Motivation}\\	[2ex]	
		\hline
		
		RQ1 & Which methods have been proposed in the literature for SSE problem? & To understand the current state-of-the-art methods for mitigating SSE problem. \\	
		
		RQ2  & How the methods mitigate SSE problem? & To enhance understanding of the theories and concepts, as well as recognizing the different between SSE reduction methods easily. It also assists with classifying the methods.  \\
		
		RQ3 & What are the key features and challenges have been obtained in the literature for SSE reduction methods? &  To identify the success factors and potential challenges that could be  encountered in using particular SSE reduction methods. \\
		
		RQ4 & Which SSE reduction methods are more frequently utilized in the literature for CBSD? & To determine the kinds of SSE reduction methods that can be apply to verify CBSD. \\
		
		RQ5 & How SSE has been tackled in CBSD?  & To enhance comprehension of the theories, concepts, success factors, and challenges.\\ 
		
		RQ6 & What are the key features and challenges have been obtained in the literature for SSE reduction methods in CBSD? & To identify how challenges could fail or limit the verification process of CBSD.\\
		
		\hline
		\hline
	\end{tabular}
\end{table*}

\subsection{Keywords, databases, and search strategy}
The keywords and search query for the RQ (1, 2, 3) were (“model-checking” AND “State space explosion problem”)  or (“model-checking” AND “the name of each mitigation methods for SSE, for example, assume-guarantee”). For RQ (4, 5, 6), we added “component-based system” to the above search strings.

To shape the keywords and search query, this work relies on “Model Checking” by E. M. Clarke \textit{et al.} \cite{clarke1999model}, and “Specification and verification of concurrent systems in CESAR” by J. Sifakis \textit{et al.} \cite{queille1982specification}, as a basis together with other studies published by those studies like \cite{ chen2010automated, clarke2011model}. One of the reasons that these books and studies have been selected is  a credit to the authors as pioneering scientists in the model-checking. 

The search queries have been executed in widely known electronic database/library resources such as ACM digital library, IEEEexplore, Science Direct, Web of Science, Springer link, Google scholar, Citeceer. 

\subsection{Inclusion and exclusion criteria}\label{inc-ex}
 In order to select the most important papers within the scope of this paper, a set of inclusion and exclusion criteria have been established. It is represented in Table \ref{tab:IncEx}.

\renewcommand{\arraystretch}{1.5}
\begin{table*}[h!]
	\caption{Inclusion and exclusion criteria }
	\label{tab:IncEx}
	\centering	
	\small
	\begin{tabular}{m{4cm} m{11 cm}}
		\hline
		\hline	 
		General inclusion criteria &    
		\begin{enumerate}
			\item Research papers, white papers, conferences, technical reports, doctorate dissertation, books, hands books.
			\item Studies the researches of well-known authors in model-checking and formal verification.
		\end{enumerate} 
		\\			
		
		Specific inclusion criteria for RQs &
		\begin{enumerate}
			\item The studies analyzes/addresses SSE in normal model-checking (RQ 1).
			\item The studies provide discussion about mitigation methods of SSE, its key features and challenges (RQ 2 and 3).
			\item The studies analyze/discuss/present verification in CBSD (RQ 4).
			\item The studies provide discussion about mitigation methods of SSE in CBSD verification, the key features and challenges (RQ 5 and 6).
			\item  The studies or tool sets that can be used to make the description more clear such as by giving an example, or illustrating other aspects or domains of the presented method.
		\end{enumerate}
		\\	
		
		Exclusion criteria & 
		\begin{enumerate}
			\item Duplicate report of the same study.
			\item  Studies to utilize the already proposed methods rather than propose a new method to tackle SSE and do not fulfill the inclusion criterion No. 5.
			\item Studies to present tool-sets based on the already proposed methods and do not fulfill the inclusion criterion No. 5.
			\item Studies that do not introduced a method for SSE reduction.
			\item Studies that utilize a SSE reduction method in a domain or programming language rather than introduce a new method. 
			\item  Studies that are more focus on deductive verification such as theorem proving. Refer to \ref{QualityAssessment} for more details.
			\item Studies that do not contain the defined keywords.
		\end{enumerate}    
		\\		
		
		\hline
		\hline
	\end{tabular}
	
\end{table*}

\subsection{Quality assessment and selected studies}\label{QualityAssessment}
After each iteration of the query, a preliminary review is carried out based on the inclusion and exclusion criteria to obtain an appropriate literature collection. Although specifying keywords and search queries, it has been observed that some results returned by the search engine are pertaining to model-checking concerning other domains or perspectives such as hardware verification. Some papers also use model-checking for specific programming languages or verify specific software.  For example, in\cite{chaki2004modular} a compositional approach has been presented and utilized for automatic verification of C programs. These papers are excluded as well.

 Some of the studies found were based on deductive methods like theorem proving. However, model-checking is an algorithmic method for deciding whether a hardware or software design meets a formal specification \cite{clarke2009model}. Due to this, studies containing methods such as theorem proving are excluded as well. For more details on difference between deductive methods and model checking, refer to \cite{silva2012case}. The emphasis of this paper is more on having insight into the SSE reduction methods and determine the methods that are suitable for verifying component-based systems.

\subsection{Information extraction}
We extract the following items from each selected paper:
\begin{enumerate}[]
	\item SSE reduction methods: the methods to mitigate SSE problem.
	\item Method's description: a brief description of SSE reduction methods.
	\item SSE reduction methods in CBSD: the frequently used SSE reduction methods in CBSD. With the focus to show which SSE reduction method  has been used frequently in (all domain of) CBSD.
	\item Key features and potential challenges: the pros and cons of  the methods that have been obtained in the literature.
	\item  Examples: Utilizing study examples/tools and illustrations of other aspects or domains of the methods to aid in the conceptual comprehension.

\end{enumerate}

\section{Main Concepts}\label{Section3}
Formal methods have great potential to verify and ensure the system’s correctness as early as possible. It removes ambiguity in the system specification and provides preciseness.  One of the well-known formal methods is \textit{Model-checking}.  It is possible to describe model-checking as a tuple:
\begin{equation}
	\mathbb{MODEL \vspace{5mm} CHECKING} = < \mathbb{M},  \mathbb{S}>
\end{equation}

Where $\mathbb{M}$ is a system model with $m$ states and $\mathbb{S}$ is the formal specification as shown in Figure \ref{Model-checking}. Let $m$ be a state of the system model, $m \in \mathbb{M}$, then model-checking searches all states $m_i$ for $1 \leq i \leq n$ in $\mathbb{M}$ and returns "Yes" if $\forall m$ satisfy $\mathbb{S}$, $(\forall m_i \in \mathbb{M}) \models \mathbb{S}$ for $1 \leq i \leq n$. Otherwise, model-checking produces counterexample(s). Counterexample(s) is a declaration that defeats the specified properties on a given system by presenting it in at least one path in the SS.

\tikzstyle{sq}=[draw,text=violet,minimum width=10pt, line width=2pt]
\tikzstyle{cr}=[draw,circle, minimum size=10pt, line width=2pt]
\tikzstyle{bcr}=[draw,circle, minimum size=5pt, line width=2pt, fill=black]
\tikzstyle{Box}=[draw,black, left color=black!20,
text=black!70]

\begin{figure}[!h]
	\centering
	\resizebox{\columnwidth}{!}{
		\begin{tikzpicture}
		\node[Box, line width=0.3mm, rounded corners, burntorange, left color=IALB!20,
		text=IADB, IADB] (center) at (0,0) {\shortstack{Model \\ Checker}};
		\node[void] (ld) at (-3.5,-1) {\shortstack{\textcolor{IADB}{$\varphi$} \\ specification}};
		\node[void] (lu) at (-3.5,1) { \shortstack{	
				\begin{tikzpicture}
				\node[cr, IADB] (b) at(0,0) {};
				\node[cr, IADB] (c) at (0.7,0) {};
				\node[cr, IADB] (a) at (-0.7,0) {};
				\node[cr, IADB] (d) at (1.4,0) {};
				\path (a) edge [->, line width=2pt, IADB] (b);
				\path (b) edge [->, line width=2pt, IADB] (c);
				\path (c) edge [->, line width=2pt, IADB] (d);
				\end{tikzpicture} \\ system model}};
		\node[void] (ru) at (3.5,1) {\shortstack{Satisfied\\ \big( \shortstack{ 	\begin{tikzpicture}
					\node[cr, IADB, fill=greengashang] (b) at(0,0) {};
					\node[cr, IADB, fill=greengashang] (c) at (0.7,0) {};
					\node[cr, IADB, fill=greengashang] (a) at (-0.7,0) {};
					\node[cr, IADB, fill=greengashang] (d) at (1.4,0) {};
					\path (a) edge [->, line width=2pt, IADB] (b);
					\path (b) edge [->, line width=2pt, IADB] (c);
					\path (c) edge [->, line width=2pt, IADB] (d);
					\end{tikzpicture} \\ witness path  } \big) }};
		\node[void] (rd) at (3.5,-1) {\shortstack{Not satisfied\\ \big(  \shortstack{ 
					\begin{tikzpicture}
					\node[cr, IADB, fill=greengashang] (b) at(0,0) {};
					\node[cr, IADB, fill=greengashang] (c) at (0.7,0) {};
					\node[cr, IADB, fill=greengashang] (a) at (-0.7,0) {};
					\node[cr, IADB, fill=red] (d) at (1.4,0) {};
					\path (a) edge [->, line width=2pt, IADB] (b);
					\path (b) edge [->, line width=2pt, IADB] (c);
					\path (c) edge [->, line width=2pt, IADB] (d);
					\end{tikzpicture}
					\\ counterexample  } \big) }};;
		\path (lu) edge [bend right=20,->, line width=2pt, IADB] (center);
		\path (ld) edge [bend left=20,->, line width=2pt, IADB] (center);
		\path (ru) edge [bend left=20,<-, line width=2pt, IADB] (center);
		\path (rd) edge [bend right=20,<-, line width=2pt, IADB] (center);
	\end{tikzpicture}
}
\caption{Model checking}
\label{Model-checking}
\end{figure}
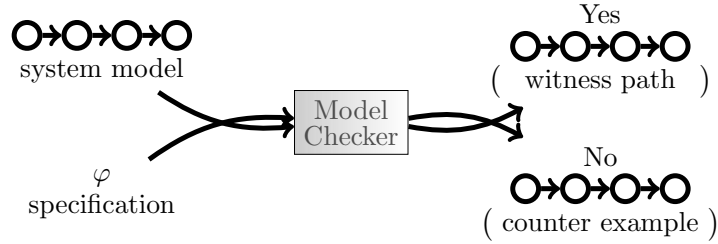

\textit{System model} $\mathbb{M}$ is a conceptual model to represent and describe a system. In model-checking, a system model is originally represented by a \textit{Kripke} structure, but can be represented as a state chart \cite{schafer2001model}, Petri net \cite{liu2016petri}, or other possible graph-like visualization of systems's SS. Using graph-like representation by individual states is one of the main representation paradigms in model-checking called \textit{explicit-state model-checking}. Figure \ref{Kripke} shows an explicit-state system model based on a \textit{Kripke} structure. The nodes in Figure \ref{Kripke}(a) (named by $X,Y,Z$) are the states of a system (or a process) modelled as nodes in the \textit{Kripke} structure. Figure~\ref{Kripke}(b) shows the SS of the modelled system in Figure \ref{Kripke}(a).

\tikzstyle{vecArrow} = [thick, decoration={markings,mark=at position
	1 with {\arrow[semithick, IADB]{open triangle 60}}},
double distance=2pt, shorten >= 5.5pt,
preaction = {decorate},
postaction = {draw,line width=3pt, white,shorten >= 4.5pt}]
\tikzstyle{innerWhite} = [semithick, white,line width=3pt, shorten >= 4.5pt]
\begin{figure}[h!]
	\centering
	\resizebox{\columnwidth}{!}{
		\begin{tikzpicture}[auto, thick, >=stealth', line width=0.5mm]
			\node(spacer) at (0,1) {};
			\node[newsuperpeers, minimum width=30pt, left color=IALB!20, text=IADB, IADB] (X) at (0,0) {$\mathbb{X}$};
			\node[newsuperpeers, minimum width=30pt, left color=IALB!20, text=IADB, IADB] (Y) at (-2,-2) {$\mathbb{Y}$};
			\node[newsuperpeers, minimum width=30pt, left color=IALB!20, text=IADB, IADB] (Z) at (2,-2) {$\mathbb{Z}$};
			\path (X) edge[<->, IADB] (Y);
			\path (Y) edge[->, IADB] (Z);
			\path (X) edge[->, IADB] (Z);
			\path (Z) edge[loop below, <-, IADB] (Z);
			
			\node[newsuperpeers, minimum width=30pt, left color=IALB!20, text=IADB, IADB] (rab) at (3.5,-3.5) {$\mathbb{X}$};
			\node[newsuperpeers, minimum width=30pt, left color=IALB!20, text=IADB, IADB] (rbc) at (4.5,-1.5) {$\mathbb{Y}$};
			\node[newsuperpeers, minimum width=30pt, left color=IALB!20, text=IADB, IADB] (rc) at (5.5,-3.5) {$\mathbb{Z}$};
			\node[newsuperpeers, minimum width=30pt, left color=IALB!20, text=IADB, IADB] (rabt) at (5.5,0.5) {$\mathbb{X}$};
			\node[newsuperpeers, minimum width=30pt, left color=IALB!20, text=IADB, IADB] (rc2) at (6.5,-1.5) {$\mathbb{Z}$};
			\node[newsuperpeers, minimum width=30pt, left color=IALB!20, text=IADB, IADB] (rc3) at (7.5,-3.5) {$\mathbb{Z}$};
			\node[void] (v1) at (2.7,-5) {};
			\node[void] (v2) at (4.2,-5) {};
			\node[void] (v3) at (5.5,-5) {};
			\node[void] (v4) at (7.5,-5) {};
			\path (rabt) edge[->, IADB] (rbc);
			\path (rabt) edge[->, IADB] (rc2);
			\path (rc2) edge[->, IADB] (rc3);
			\path (rbc) edge[->, IADB] (rc);
			\path (rbc) edge[->, IADB] (rab);
			\draw[-, line width=0.5mm, IADB, dotted] (rab) to (v1);
			\draw[-, line width=0.5mm, IADB, dotted] (rab) to (v2);
			\draw[-, line width=0.5mm, IADB, dotted] (rc) to (v3);
			\draw[-, line width=0.5mm, IADB, dotted] (rc3) to (v4);
			
			\node[legend_general] at (0,-6) {(a)};
			\node[legend_general] at (6,-6) {(b)};
		\end{tikzpicture}
	}
	\caption{A kripke structure and its computation tree \textmd{\cite{clarke1999model}}}
	\label{Kripke}
\end{figure}

Another representation paradigm is the \textit{implicit model-checking}. In this model, states checking are not individually represented, but a quantified propositional logic formula is used to represent the state space graph.

The formal specification in model-checking is represented by propositional temporal logic which is a kind of logic to specify and reason over the ongoing properties of the system being modelled in terms of time. There are two types of operations that temporal logic supports: The first are logical operators such as $\neg, \vee, \wedge$, and the second refers to modalities like \textbf{F}inally, \textbf{U}ntil, \textbf{G}lobally. The set of operators which expresses properties in only a single future position for every point in running time is called \textit{linear temporal logic (LTL)}.

In Figure \ref{LTLModelchecking}, three sequences of events of a system (or a process) is shown with two properties, $p$ and $q$.  LTL checks only one sequence of event(s) in a single run and does not switch to another run while checking. For example, in the first sequence of events in Figure \ref{LTLModelchecking}, $F_p$ (finally properties $p$ happens) will be checked. The light blue state, $p$, will finally happen after the other states (indicated by dark blue) is checked. For the second sequence of events in Figure \ref{LTLModelchecking}, $p\ U\ q$ ($p$ can continue to happen until $q$ happens) will be checked. The light blue and green states are $p$ and $q$ accordingly. For the third sequence of events, $Gp$ (globally $p$ may happen in the future) among all of the states in the sequence, $p$ may happen. Due to this, all the states are in light blue.  

\begin{figure}[]
	\centering
	\resizebox{\columnwidth}{!}{
		\begin{tikzpicture}[auto, thick, >=stealth', line width=0.5mm]
			\node(spacer) at (0,0.5) {};
			\node[superpeersgry, left color=IADB, IADB] (00) at (0,0) {};
			\node[superpeersgry, left color=IADB, IADB] (01) at (1,0) {};
			\node[superpeersgry, left color=IADB, IADB] (02) at (2,0) {};
			\node[superpeersgry, left color=IADB, IADB] (03) at (3,0) {};
			\node[superpeersgry, left color=IADB, IADB] (04) at (4,0) {};
			\node[superpeersgry, left color=IADB, IADB] (05) at (5,0) {};
			\node[newsuperpeers, left color=IALB!20, IADB] (06) at (6,0) {};
			\node[superpeersgry, left color=IADB, IADB] (07) at (7,0) {};
			\node[void] (08) at (8,0) {};
			\path (00) edge[IADB] (01);
			\path (01) edge[IADB]  (02);
			\path (02) edge[IADB]  (03);
			\path (03) edge[IADB]  (04);
			\path (04) edge[IADB]  (05);
			\path (05) edge[IADB]  (06);
			\path (06) edge[IADB]  (07);
			\path (07) edge[->, IADB] (08);
			\node[legend_general] at (4,1) {\large Finally \textcolor{black}{p}};
			\node[legend_general] at (4,-1) { $F \textcolor{black}{_P}$};	
			\node[newsuperpeers, left color=IALB!20, IADB] (10) at (0,-3.5) {};
			\node[newsuperpeers, left color=IALB!20, IADB] (11) at (1,-3.5) {};
			\node[newsuperpeers, left color=IALB!20, IADB] (12) at (2,-3.5) {};
			\node[newsuperpeers, left color=IALB!20, IADB] (13) at (3,-3.5) {};
			\node[newsuperpeers, left color=IALB!20, IADB] (14) at (4,-3.5) {};
			\node[newsuperpeers, left color=IALB!20, IADB] (15) at (5,-3.5) {};
			\node[newsuperpeersvio, left color=greengashang, IADB] (16) at (6,-3.5) {};
			\node[superpeersgry, left color=IADB, IADB] (17) at (7,-3.5) {};
			\node[void] (18) at (8,-3.5) {};
			\path (10) edge[IADB] (11);
			\path (11) edge[IADB] (12);
			\path (12) edge[IADB] (13);
			\path (13) edge[IADB] (14);
			\path (14) edge[IADB] (15);
			\path (15) edge[IADB] (16);
			\path (16) edge[IADB] (17);
			\path (17) edge[->, IADB] (18);
			\node[legend_general] at (4,-2.5) {\large \textcolor{black}{p} \textcolor{black}{Until} \textcolor{greengashang}{q}};
			\node[legend_general] at (4,-4.5) { \textcolor{black}{p} \textcolor{black}{U} \textcolor{greengashang}{q}};	
			\node[newsuperpeers, left color=IALB!20, IADB] (20) at (0,-7) {};
			\node[newsuperpeers, left color=IALB!20, IADB] (21) at (1,-7) {};
			\node[newsuperpeers, left color=IALB!20, IADB] (22) at (2,-7) {};
			\node[newsuperpeers, left color=IALB!20, IADB] (23) at (3,-7) {};
			\node[newsuperpeers, left color=IALB!20, IADB] (24) at (4,-7) {};
			\node[newsuperpeers, left color=IALB!20, IADB] (25) at (5,-7) {};
			\node[newsuperpeers, left color=IALB!20, IADB] (26) at (6,-7) {};
			\node[newsuperpeers, left color=IALB!20, IADB] (27) at (7,-7) {};
			\node[void] (28) at (8,-7) {};
			\path (20) edge[IADB] (21);
			\path (21) edge[IADB] (22);
			\path (22) edge[IADB] (23);
			\path (23) edge[IADB] (24);
			\path (24) edge[IADB] (25);
			\path (25) edge[IADB] (26);
			\path (26) edge[IADB] (27);
			\path (27) edge[->, IADB] (28);
			\node[legend_general] at (4,-6) {\large Globally \textcolor{black}{p}};
			\node[legend_general] at (4,-8) { $G \textcolor{black}{_P}$};	
		\end{tikzpicture}	
	}
	\caption{Example of LTL operators \textmd{\cite{clarke2011model}}}
	\label{LTLModelchecking}
\end{figure}

\begin{figure*}[h]
	\centering
	\begin{tikzpicture}[shorten >=1pt,auto,node distance=2.0cm,
		line width=0.3mm]
		\node[newsuperpeers, left color=IALB!20, IADB] (r00) at (0,0) {};
		\node[newsuperpeers, left color=IALB!20, IADB] (r10) at (-.5,-1) {};
		\node[newsuperpeers, left color=IALB!20, IADB] (r11) at (.5,-1) {};
		\node[newsuperpeers, left color=IALB!20, IADB] (r20) at (-1,-2) {};
		\node[newsuperpeers, left color=IALB!20, IADB] (r21) at (0,-2) {};
		\node[newsuperpeers, left color=IALB!20, IADB] (r22) at (1,-2) {};
		\node[newsuperpeers, left color=IALB!20, IADB] (r30) at (-1.5,-3) {};
		\node[newsuperpeers, left color=IALB!20, IADB] (r31) at (-.5,-3) {};
		\node[newsuperpeers, left color=IALB!20, IADB] (r32) at (.5,-3) {};
		\node[newsuperpeers, left color=IALB!20, IADB] (r33) at (1.5,-3) {};
		\node[newsuperpeers, left color=IALB!20, IADB] (r40) at (-2,-4) {};
		\node[newsuperpeers, left color=IALB!20, IADB] (r41) at (-1,-4) {};
		\node[newsuperpeers, left color=IALB!20, IADB] (r42) at (0,-4) {};
		\node[newsuperpeers, left color=IALB!20, IADB] (r43) at (1,-4) {};
		\node[newsuperpeers, left color=IALB!20, IADB] (r44) at (2,-4) {};
		\node[newsuperpeers, left color=IALB!20, IADB] (r50) at (-2.5,-5) {};
		\node[newsuperpeers, left color=IALB!20, IADB] (r51) at (-1.5,-5) {};
		\node[newsuperpeers, left color=IALB!20, IADB] (r52) at (-.5,-5) {};
		\node[newsuperpeers, left color=IALB!20, IADB] (r53) at (.5,-5) {};
		\node[newsuperpeers, left color=IALB!20, IADB] (r54) at (1.5,-5) {};
		\node[newsuperpeers, left color=IALB!20, IADB] (r55) at (2.5,-5) {};
		\path (r00) edge[IADB] (r10);
		\path (r00) edge[IADB] (r11);
		\path (r10) edge[IADB] (r20);
		\path (r10) edge[IADB] (r21);
		\path (r11) edge[IADB] (r22);
		\path (r20) edge[IADB] (r30);
		\path (r21) edge[IADB] (r31);
		\path (r22) edge[IADB] (r32);
		\path (r22) edge[IADB] (r33);
		\path (r30) edge[IADB] (r40);
		\path (r30) edge[IADB] (r41);
		\path (r31) edge[IADB] (r41);
		\path (r31) edge[IADB] (r42);
		\path (r32) edge[IADB] (r42);
		\path (r33) edge[IADB] (r43);
		\path (r33) edge[IADB] (r44);
		\path (r40) edge[IADB] (r50);
		\path (r41) edge[IADB] (r51);
		\path (r42) edge[IADB] (r52);
		\path (r42) edge[IADB] (r53);
		\path (r43) edge[IADB] (r54);
		\path (r44) edge[IADB] (r55);
		\node[] (00) at (0,-6) {(a) $AG_p$};
		
		\node(spacer) at (4,0.5) {};
		\node[newsuperpeers, left color=IALB!20, IADB] (00) at (7,0) {};
		\node[superpeersgry, left color=IADB, IADB] (10) at (6.5,-1) {};
		\node[newsuperpeers, left color=IALB!20, IADB] (11) at (7.5,-1) {};
		\node[superpeersgry, left color=IADB, IADB] (20) at (6,-2) {};
		\node[superpeersgry, left color=IADB, IADB] (21) at (7,-2) {};
		\node[newsuperpeers, left color=IALB!20, IADB] (22) at (8,-2) {};
		\node[superpeersgry, left color=IADB, IADB] (30) at (5.5,-3) {};
		\node[superpeersgry, left color=IADB, IADB] (31) at (6.5,-3) {};
		\node[newsuperpeers, left color=IALB!20, IADB] (32) at (7.5,-3) {};
		\node[superpeersgry, left color=IADB, IADB] (33) at (8.5,-3) {};
		\node[superpeersgry, left color=IADB, IADB] (40) at (5,-4) {};
		\node[superpeersgry, left color=IADB, IADB] (41) at (6,-4) {};
		\node[newsuperpeers, left color=IALB!20, IADB] (42) at (7,-4) {};
		\node[superpeersgry, left color=IADB, IADB] (43) at (8,-4) {};
		\node[superpeersgry, left color=IADB, IADB] (44) at (9,-4) {};
		\node[superpeersgry, left color=IADB, IADB] (50) at (4.5,-5) {};
		\node[superpeersgry, left color=IADB, IADB] (51) at (5.5,-5) {};
		\node[superpeersgry, left color=IADB, IADB] (52) at (6.5,-5) {};
		\node[newsuperpeers, left color=IALB!20, IADB] (53) at (7.5,-5) {};
		\node[superpeersgry, left color=IADB, IADB] (54) at (8.5,-5) {};
		\node[superpeersgry, left color=IADB, IADB] (55) at (9.5,-5) {};
		\path (00) edge[IADB] (10);
		\path (00) edge[IADB] (11);
		\path (10) edge[IADB] (20);
		\path (10) edge[IADB] (21);
		\path (11) edge[IADB] (22);
		\path (20) edge[IADB] (30);
		\path (21) edge[IADB] (31);
		\path (22) edge[IADB] (32);
		\path (22) edge[IADB] (33);
		\path (30) edge[IADB] (40);
		\path (30) edge[IADB] (41);
		\path (31) edge[IADB] (41);
		\path (31) edge[IADB] (42);
		\path (32) edge[IADB] (42);
		\path (33) edge[IADB] (43);
		\path (33) edge[IADB] (44);
		\path (40) edge[IADB] (50);
		\path (41) edge[IADB] (51);
		\path (42) edge[IADB] (52);
		\path (42) edge[IADB] (53);
		\path (43) edge[IADB] (54);
		\path (44) edge[IADB] (55);
		\node[] (00) at (7,-6) {(b) $EG_p$};
	\end{tikzpicture}
	\caption{Example of CTL operators\textmd{\cite{clarke2011model}}}
	\label{CTLModelchecking}
\end{figure*}

On the other hand, there is another type of temporal logic called \textit{computational tree logic (CTL)} that is able to check all possible paths in a single run. In Figure \ref{CTLModelchecking}(a), the light blue nodes are all possible paths, which CTL can switch  between any of them during one single  run. It checks by operator $AG_p$ (in  all paths, the property $p$ may happen). CTL operators can support one sequence of events in one run as well.  For example, in Figure \ref{CTLModelchecking}(b), the light blue path is one sequence of events. This sequence will be checked by operator $EG_p$ (eventually in one of the paths, the property $p$ may happen) to indicate whether $p$ eventually happens or not.

\textit{System property} is defined by temporal logic. Some common system properties are \textit{reachability properties} (some particular position in a system model that can be met), \textit{Safety properties} (under particular circumstances, an event will not happen in any way, like without the key a car won't start.), \textit{liveness} (under certain circumstances, some event will eventually happen, like if we press the button of an elevator, it is bound to arrive ultimately.), \textit{fairness} (under certain circumstances an event will or will not happen infinitely often, like the gate will be raised infinitely often.) \cite{clarke1999model}.

\textit{State space explosion} is the most critical problem restricting model-checking. To define SSE, let a given system contains $n$ processes and each process has $m$ states (same states). The size of SS of these $n$ processes might be $m^n$. Then, the amount of state space of a given system may increase exponentially with the size of its states (processes) and consequently exceed the memory capacity of the system. 

\section{Results on SSE reduction methods}\label{Section4}
This section elaborates the results of SSE reduction methods that have been found in  the research literature. The results are represented as a classification of SSE reduction methods. The classification has five dimensions summarized as the following:

\begin{enumerate}
	\item \textit{Memory handling.} Identifies the methods that  are directly engaged in memory expansion and management.
	\item \textit{Heuristics and probabilistic reasoning.} Identifies the methods that are able to find an approximation of the exact solution. They are the fastest way to find a close solution when the exact answer cannot be computed.
	\item \textit{Scaling down the state space.} Identifies the methods that try to reduce the size of states to be stored in the memory. The reduction can be based on compression, symmetry, and similarity omission, using binary decision diagram (BDD), or hash table.
	\item \textit{Bottom-up approach.} Identifies the methods that start verification as early as the whole state space is constructed.
	\item \textit{Divide-and-conquer approach.} Identifies the methods which decompose the state space into small parts and address each small  part separately.
\end{enumerate}

\textit{SSE} problem is a bottleneck in model-checking. The amount of a system's SS (even a finite system) strongly depends on its components and is prone to increase in size exponentially.  Consequently, it easily exceeds the computer's memory capacity and limits the size that can be verified by a model checker. Therefore, memory becomes  the main concern for SSE reduction methods. Memory concern can be addressed from several aspects, for example, expanding memory capacity, reducing the states that need to be stored in memory, released memory from the redundant and repeated states, etc. Generally, a method cannot cover all the aspects, some methods find and omit redundancies, while others may focus on memory expansion. Therefore, this work simply classifies these methods according to the aspects that each method can cover.

A discussion about the classification is provided in detail in the first subsection and an illustration of it is represented in Figure \ref{something}. Each class is divided into multiple sub-classes which is in accordance with the current methods for tackling SSE problem. The RQs (1, 2) are answered in this section. The second part of this section pertains to RQ 3. The characteristics, key features, and challenges of SSE reduction methods have been summarized in tables.

\definecolor{blk40}{HTML}{0066A1}
\definecolor{blk30}{HTML}{019AD3}
\definecolor{blk20}{HTML}{2fc5fe}
\definecolor{blk3}{HTML}{97e2fe}
\begin{figure*}[!t]
	\begin{minipage}{\textwidth}
		\centering
		\resizebox{\textwidth}{!}{
			\begin{tikzpicture}[ every annotation/.style = {draw, fill = white}]
				\path[mindmap,concept color=blk40,text=white,
				every node/.style={concept},
				root/.style    = {concept color=blk30,font=\bfseries,text width=6em},
				level 1 concept/.append style={color=black, font=\bfseries,text width=7em, level distance=14em,inner sep=0pt},
				level 2 concept/.append style={level distance=9em},
				]
				node[root] {SSE\\Reduction\\Methods} [clockwise from=0]
				child[concept color=blk20, sibling angle=45] {node {Compositional Verification} [clockwise from=54]
					child[concept color=blk3] { node {Interface Rule}}
					child[concept color=blk3, sibling angle=36] { node {Assume Guarantee Reasoning}}
					child[concept color=blk3, sibling angle=36] { node {Lazy Parallel Compisition}}
					child[concept color=blk3, sibling angle=36] { node {Partial Transition Relation}}
				}
				child[concept color=blk20, , sibling angle=45] {node[concept] {Memory Handelling}[clockwise from=320]
					child[concept color=blk3] { node[concept]{Expanding Memory}}
					child[concept color=blk3] { node[concept]{Garbage Collection Reduction}}
				}
				child[concept color=blk20, , sibling angle=65] {node[concept]{Bottom-Up Methods}[clockwise from=270]
					child[concept color=blk3] { node[concept]{On-the-fly Verification}}
					child[concept color=blk3] { node[concept]{Incremental Verification}}
				}
				child[concept color=blk20, , sibling angle=60] {node[concept]{Heuristic and Probabilistic Methods}[clockwise from=234]
					child[concept color=blk3] { node[concept] {Genetic Algorithm \& Ant Colony}}
					child[concept color=blk3, sibling angle=36] { node[concept] {Random Walk }}
					child[concept color=blk3, sibling angle=36] { node[concept] {Machine Learning}}
					child[concept color=blk3, sibling angle=36] { node[concept] {Bloom Filter}}
				}
				child[concept color=blk20, , sibling angle=67.5 ] {node[concept] {Scaling down the State Space}[counterclockwise from=0]
					child[concept color=blk3] { node[concept]{Abstraction}}
					child[concept color=blk3, sibling angle=180] { node[concept]{Bounded Model Checking}}
					child[concept color=blk3, sibling angle=65] { node[concept]{Partial Order Reduction}}
					child[concept color=blk3, sibling angle=30] { node[concept]{Symbolic Model Checking}}
					child[concept color=blk3, sibling angle=12] { node[concept]{Symmetry Reduction}}
				}
				; 			
			\end{tikzpicture}
		}
		\caption{An overview of SSE reduction methods }
		\label{something}
	\end{minipage}
\end{figure*}
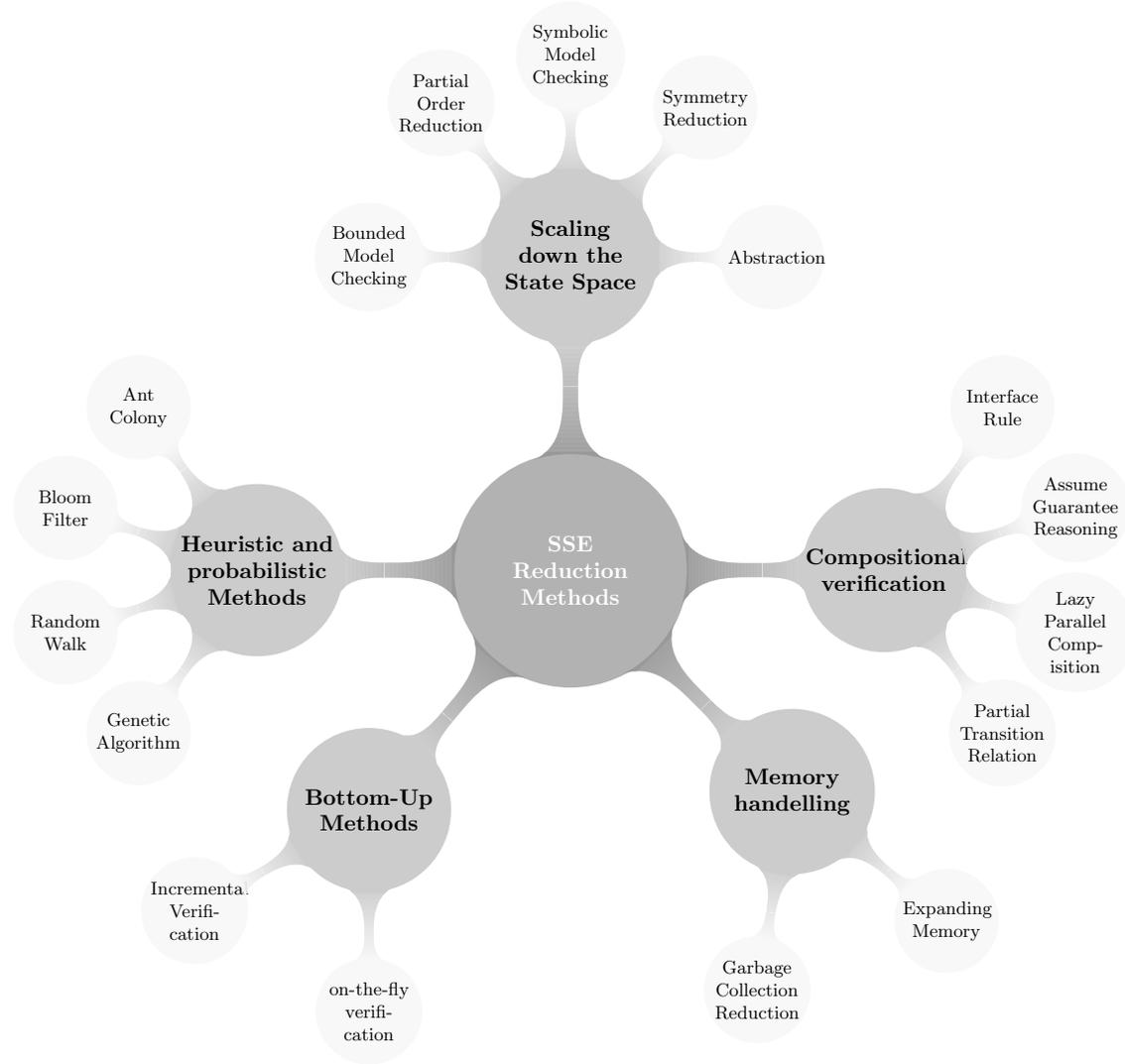

\subsection{Classification}
 The classified SSE reduction methods are described below based on evaluating the selected studies:

\subsubsection{Memory handling}
 As previously described, memory is the main concern of SSE reduction methods  that can be addressed from several aspects. Some of them are  specifically involved in memory and are  contradictory to the other methods like compressing the state space. These kinds of methods are based on the following principles: 1- proper memory management to increase performance. Memory management is a process of controlling and incorporating programs by using a sufficient methodology to fragment, allocate, monitor, and release memory. 2- increase memory capacity to provide more space for data storage. It can be achieved by expanding the external memory. The aforementioned two principles can be achieved through the following:

\paragraph{Expanding external memory}
The idea of using an external memory with  a proper algorithm when the RAM cannot handle all data, might be one of the solutions to overcome SSE problem.  External memory is able to provide a much larger space. Currently, the capacity of magnetic disks is increasing enormously at  a relative cost. This fact motivates researchers to utilize external memories in model-checking. Due to the fact that external memory cannot be accessed rapidly like internal memory, providing an efficient external memory algorithm is the main concern in using this method. The algorithm must organize disk access carefully and precisely. The efficiency of the algorithm is determined via the amount of I/Os. In other words, between the amount of I/Os and time efficiency, there is a relationship in the sense that the time efficiency will be improved if the I/O actions are reduced.

S. Edelkamp \textit{et al.} \cite{lamborn2008layered} proposed layered duplicate detection to improve duplicate elimination in external memory model-checking. This approach determines which states, while searching in the state space, should be stored in the RAM, and which of them should be stored on disk. As a result, it increases the efficiency of the run time and decreases the amount of disk storage. 

W. Lijun \textit{et al.} \cite{wu2015efficient} proposed an I/O (input/output) efficient methodology to provide a model checker based on extending memory. Their methodology is generally based on  nested depth-first (NDF). By combining the following three methods the authors have achieved a significant improvement in time efficiency of I/Os. The first method is sorting a hash table in linear time. In this method, the already visited states sorted in a hash table will be merged into a hash table which is saved and sorted into the external memory. The second method is detecting duplicates in a cache. Finally, the third method refers to the managing of the dynamic path.  This methodology gives performance guarantees for I/O efficiency.

\paragraph{Garbage collection reduction}
Garbage collection reduction is a memory management policy  that is inspired by utilizing garbage collection in real-life software systems to improve model-checking methods. It deletes information about already visited-states and reclaims allocated memory while model-checking is performed.  On the contrary, when garbage collection and memory reclaiming for idle memory is not utilized, the state space may grow and limit the verification. Garbage collection has some advantages, for example, it does not suffer from inefficient memory fragmentation and complex pointer analysis. 

One of the classic collection algorithms is by Mark and Sweep. In this algorithm, a state can be marked as garbage when no more transition to it is available \cite{lerda2001addressing}. In other words, it is based on reachability \cite{iosif2000using} and uses graph-search algorithms like depth-first search to indicate any state that must be marked as garbage. 

The other collection algorithm is reference counting collection \cite{appel2004modern}. It discovers the garbage directly by monitoring and counting the pointers that point to each state. Disabling to reclaim the cycles of garbage is the major difficulty of this algorithm \cite{iosif2000using}. Additional algorithm for garbage collections is based on finding usability \cite{might2007model}. The most widely used garbage collections are Java based programs \cite{lengauer2014taming}.

\subsubsection{Heuristic and probabilistic reasoning}
In many problems that do not have an exact solution, the hope is to have at least an approximate answer. In this case, probabilistic reasoning can handle the situation. It is able to find   approximate solution faster than other methods. The solution may not be optimum, however, is still valuable because it helps us in cases that the exact solution could not be achieved. Utilizing this kind of   method is a possible way to reduce the model-checking effort. The rest of this section introduces a few SSE reduction methods based on probabilistic reasoning.

\paragraph{Genetic algorithms}
A genetic algorithm (GA) in computer science is a meta-heuristic optimizer that is based on population (a set of chromosomes) and encouraged by biological evolution. GA is a subset of a larger category called evolutionary algorithms (EA) and has been used in model-checking \cite{zheng2018genetic}. In some cases, an existing exact method (the methods which  try to find an exact solution, not an approximate solution) can fail to detect  an exact or complete solution for a given problem or there is no solution with lower complexity to them. Thus, it may be sufficient to find solutions approximately or to provide faster coverage approaching solutions by using heuristics such as GA. In analyzing the use of GA or any EAs in model-checking, one must consider to target finding any solution (any error) not only the optimum solution. In addition, every reachable state of the entire system must also be checked. 

P. Godefroid and S. Khurshid investigated the utilization of GA in order to explore very large state space for finding errors such as deadlock and assertions violations \cite{godefroid2002exploring}. They combined model-checking and GA to guide searching during the verification problem of a concurrent reactive system. When there are more than one enabled transitions in the current state, GA tries to explore them and find the transition which is the most fitted to be selected by using a fitness function.

R. Yousefian \textit{et al.} \cite{yousefian2014heuristic} explored the use of GA for model-checking of graph transformation systems. In their work, an incomplete SS is created instead of creating an entire SS to detect deadlock. Model checker only checks paths with a low outgoing transition. Another article found in the conducted search \cite{ma2019probabilistic} presented a mixed way of genetic algorithm and assume-guarantee reasoning (this type of model-checking is introduced later in this paper) to mitigate SSE. 

To enhance the results of using EAs in model-checking, other state-of-the-art algorithms like the Imperialist competitive algorithm \cite{atashpaz2007imperialist}, Grey-wolf optimization algorithm \cite{mirjalili2014grey}, or Raccoon optimization algorithm \cite{koohi2018raccoon} may help.

\paragraph{Random walk}
Random walk defines a path that includes a sequence of random steps to find errors in model-checking \cite{pelanek2005enhancing}. For a certain type of graph like the Markov chain, random walk is able to decide reachability and predict error traces by polynomial algorithms. The complexity of the algorithms is not better than some other methods for alleviating SSE problem, instead, it is worst. However, some advantages exist and are of interest to researchers  to use in the verification processes. For instance, its need for memory space is minimal which is an advantage. Secondly, parallel random walk is easy to implement and reduces execution time. Despite the advantages, it does not guarantee the exploration of all global states.

A tool for randomization search in SS has been proposed by D. Owen \textit{et al.} \cite{owen2003advantages} called LURCH. The researchers compared the tool with other tools and concluded that LURCH cannot be preferable as much as the others that have a complete search feature. Nevertheless, futuristics and random search can be useful to some system models that are massive and cannot be explored completely. A random walk based heuristic algorithms  are presented in \cite{sivaraj2003random, 5368102}

\paragraph{Bloom filter}
The two main schemes for probabilistic verification are hash compaction and bit state hashing, which utilizes the data structure of Bloom filter. Bloom filter is an explicit and probabilistic method for verification activities. It stores compressed values in a hash table rather than storing full state descriptors. During the verification process,  the states with a non-zero probability will be deleted. Therefore, some reachable states are never checked during the verification process and may result in false-positive outcomes.

An improved probabilistic method based on this method has been proposed by U. Stern and  D. L. Dill in \cite{stern1995improved}. The researchers reduce the probability of deleting states by using a specific hashing design. The design requires a lower number of probes required in the hash table. Another work of them for   the probabilistic method is presented in \cite{stern1996new}.

P. C. Dillinger and P. Manolios \cite{dillinger2004bloom} proposed a method based on Bloom filter which is more accurate that shows the Bloom filter can play an important role in model-checking.

\paragraph{Ant colony}
 Ant colony method is another probabilistic method to optimize the problems which is mostly used to find the optimum paths through graph-like models. It can be also applied in model-checking and verification. L. M. Duarte \textit{et al.} \cite{duarte2010model} combined model-checking and the ant colony method to solve the traveling salesman problem.

\paragraph{Machine Learning}
Machine learning (ML) is another method used to train the data set of a verification process. The data set could be the system modeled through graph-like visualization such as \textit{Kripke} structures, CTL or LTL formulas, and the results obtained from a model-checking tool. Subsequently, ML trains the data set to predict the results. The related articles based on the conducted search in this paper for machine learning in model-checking are \cite{zhu2019approximate} and \cite{zhu2019ltl}.

Another kind of heuristic model-checking method to mitigate SSE found through the search result is the continuous-time Markov chain \cite{ACM2, ACM3}. A hybrid metaheuristic approach is also presented in \cite{rezaee2020hybrid}. Heuristic model checking could be utilized by other SSE reduction methods to get more optimum results. For example in \cite{liu2020heuristics} a heuristics-based incremental model checking at runtime has been introduced. 

\subsubsection{Scaling down the state space}
Scaling down the size of SS to be checked by model checkers is another way to alleviate SSE problem. One can begin by representing the SS in another way (implicitly) which consumes less memory, like symbolic representation instead of defining them in the original shape (explicitly), which truly compresses the SS. In the comparison part of this section, in Table \ref{tab:implicit}, the implicit and explicit methods have been indicated. Furthermore, capturing a critical part of the system \cite{ACM} and ignoring irrelevant or useless variables and information in a system leads to a decrease in the size of the SS. Additionally, discovering duplicate states and avoiding regenerating them is another way of reducing SS.

The following briefly discusses some methods for alleviating SSE problem for  the state space scaling down class.

\paragraph{Symbolic model checking}
This method is utilized to compress the SS of a system by symbolically (implicitly) representing the SS. It considers a large number of states in a single step and represents them as formulas instead of enumerating them one at a time. As a result, representing them in such a way, reduces the size of the SS. It was introduced by J. R. Burch \textit{et al.} \cite{burch1992symbolic} based on Bryant's binary decision diagram (BDD) for Mu-Calculus. BDD is a data structure used to canonically represent a Boolean formula that is essentially compressed even more than other data structures, and Mu-Calculus falls into a kind of logic called modal logic (a type of logic that is able to express  modalities like a possibility and impossibility) which is able to define the properties in terms of graph-like patterns.

The methods have been used successfully for many problems such as to derive efficient decision procedures for CTL, and satisfiability of LTL. For example, a verification tool-set called ITS-tools by Y. Thierry-Mieg \cite{thierry2015symbolic} has been developed based on symbolic model-checking which supports reachability property and two kinds of temporal logic, CTL and LTL of the concurrent specification. Symbolic model-checking also has been used in a diverse range of systems like distributed control systems \cite{guellouz2018designing}.

However, the BDD that is a substantial part of symbolic model-checking extremely relies upon the variable's ordering which limits the use of symbolic model-checking. To illustrate it more precisely, we use the following example:

Let two orders of a Boolean functions with 6 variables \cite{kissmann2014bdd, bryant1992symbolic}:
$$1- (a.b) + (c.d) + (e.f)$$ $$2- (a.d) + (b.e) + (c.f)$$ 

\begin{figure}[h!]
	\centering
	\resizebox{\columnwidth}{!}{
		\begin{tikzpicture}[auto, thick]
			\node[superpeers, left color=IALB!20, text=IADB, IADB] (a1) at (0,0) {1};
			\node[superpeers, left color=IALB!20, text=IADB, IADB] (a2) at (1,-1) {2};
			\node[superpeers, left color=IALB!20, text=IADB, IADB] (a3) at (0,-2) {3};
			\node[superpeers, left color=IALB!20, text=IADB, IADB] (a4) at (1,-3) {4};
			\node[superpeers, left color=IALB!20, text=IADB, IADB] (a5) at (0,-4) {5};
			\node[superpeers, left color=IALB!20, text=IADB, IADB] (a6) at (1,-5) {6};
			\node[peers, bottom color=IALB!20, text=IADB, IADB] (al1) at (0,-6) {0};
			\node[peers, bottom color=IALB!20, text=IADB, IADB] (al2) at (2,-6) {1};
			\path (a1) edge[IADB] (a2);
			\path (a1) edge[IADB] (a3);
			\path (a2) edge[IADB] (a3);
			\path (a3) edge[IADB] (a4);
			\path (a3) edge[IADB] (a5);
			\path (a4) edge[IADB] (a5);
			\path (a5) edge[IADB] (a6);
			\path (a5) edge[IADB] (al1);
			\path (a6) edge[IADB] (al2);
			\path (a6) edge[IADB] (al1);
			\path (a4) edge[IADB] (al2);
			\path (a2) edge[IADB] (al2);
			\node[superpeers, left color=IALB!20, text=IADB, IADB] (b1) at (6,0) {1};
			\node[superpeers, left color=IALB!20, text=IADB, IADB] (b2r) at (8,-1) {2};
			\node[superpeers, left color=IALB!20, text=IADB, IADB] (b2l) at (4,-1) {2};
			\node[superpeers, left color=IALB!20, text=IADB, IADB] (b3ll) at (3,-2) {3};
			\node[superpeers, left color=IALB!20, text=IADB, IADB] (b3lr) at (5,-2) {3};
			\node[superpeers, left color=IALB!20, text=IADB, IADB] (b3rl) at (7,-2) {3};
			\node[superpeers, left color=IALB!20, text=IADB, IADB] (b3rr) at (9,-2) {3};
			\node[superpeers, left color=IALB!20, text=IADB, IADB] (b41) at (10,-3) {4};
			\node[superpeers, left color=IALB!20, text=IADB, IADB] (b42) at (8.5,-3) {4};
			\node[superpeers, left color=IALB!20, text=IADB, IADB] (b43) at (7.5,-3) {4};
			\node[superpeers, left color=IALB!20, text=IADB, IADB] (b44) at (5.5,-3) {4};
			\node[superpeers, left color=IALB!20, text=IADB, IADB] (b51) at (7.7,-4) {5};
			\node[superpeers, left color=IALB!20, text=IADB, IADB] (b52) at (4,-4) {5};
			\node[superpeers, left color=IALB!20, text=IADB, IADB] (b6) at (5.5,-5) {6};
			\node[peers, bottom color=IALB!20, text=IADB, IADB] (bl0) at (3,-6) {0};
			\node[peers, bottom color=IALB!20, text=IADB, IADB] (bl1) at (10,-6) {1};
			\path (b1) edge[IADB] (b2r);
			\path (b1) edge[IADB] (b2l);
			\path (b2l) edge[IADB] (b3ll);
			\path (b2l) edge[IADB] (b3lr);
			\path (b2r) edge[IADB] (b3rr);
			\path (b2r) edge[IADB] (b3rl);
			\path (b3ll) edge[IADB] (bl0);
			\path (b3ll) edge[IADB] (b6);
			\path (b3lr) edge[IADB] (b51);
			\path (b3lr) edge[IADB] (b52);
			\path (b3rl) edge[IADB] (b44);
			\path (b3rl) edge[IADB] (b42);
			\path (b3rr) edge[IADB] (b41);
			\path (b3rr) edge[IADB] (b43);
			\path (b44) edge[IADB] (bl0);
			\path (b44) edge[IADB] (bl1);
			\path (b43) edge[IADB] (b52);
			\path (b43) edge[IADB] (bl1);
			\path (b42) edge[IADB] (b6);
			\path (b42) edge[IADB] (bl1);
			\path (b41) edge[IADB] (b51);
			\path (b41) edge[IADB] (bl1);
			\path (b51) edge[IADB] (b6);
			\path (b51) edge[IADB] (bl1);
			\path (b52) edge[IADB] (bl0);
			\path (b52) edge[IADB] (bl1);
			\path (b6) edge[IADB] (bl0);
			\path (b6) edge[IADB] (bl1);
			
			\node[legend_general] at (0,1) {\small{\textsc{$a\times b + c \times d + e \times f$}}};
			\node[legend_general] at (6,1) {\small{\textsc{$a \times d + b \times e + c \times f$}}};
		\end{tikzpicture}
	}
	\caption{Ordering Dependency \cite{kissmann2014bdd}}
	\label{fig1}
\end{figure}
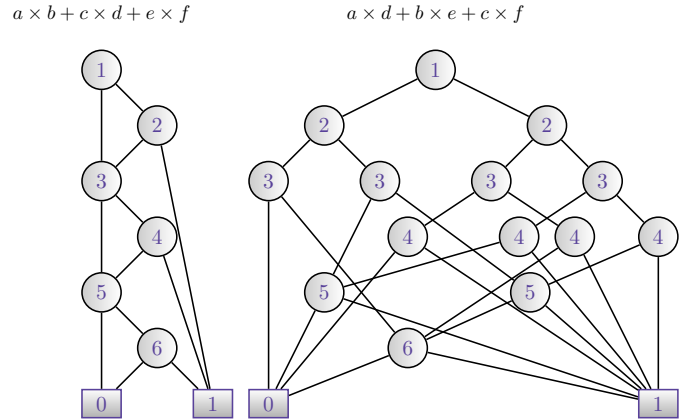

These two Boolean functions have the same number of variables, but they are different in order. The constructed BDD for both, as indicated in Figure \ref{fig1}, are not the same because BDD is sensitive to ordering. For the first function, the BDD has fewer nodes while for the second function it has more nodes.

Thus, reduced ordered binary decision diagrams (ROBDD) are used to reduce decision graphs and provide a more concise canonical representation for Boolean propositions. An improved variable ordering of BDD was introduced by  P. Prasad \textit{et al.} \cite{prasad2006binary} that is based on graph topology. They demonstrated that using a graph representation of a given Boolean function and computing the shortest path among the variables can improve ROBDD. Further work to improve ROBDD was presented by P. K. Sharma \textit{et al.} \cite{sharma2014improved} to get the most optimum size of ROBDD. However, computing an optimum order for ROBDD generally falls into the NP-Complete problem category proved by B. Bolling \textit{et al.} in \cite{bollig1996improving} and B. Bolling in \cite{bollig2014width}. A parallel version of symbolic model-checking was also presented to improve the sequential version \cite{ouni2017parallel} and \cite{ouni2019towards}.

\begin{figure}[h!]
	\centering
	\resizebox{\columnwidth}{!}{
		\begin{tikzpicture}[auto, thick]
			\node[void] (a0) at (0.3,-1) {};
			\node[void] (v2) at (0,-1) {};
			\node[superpeersgry, left color=IADB, IADB] (a1) at (6,-1) {};
			\node[superpeersgry, left color=IADB, IADB] (a21) at (7,-2) {};
			\node[superpeersgry, left color=IADB, IADB] (a22) at (5,-2) {};
			\node[superpeersgry, left color=IADB, IADB] (a31) at (8,-3) {};
			\node[superpeersgry, left color=IADB, IADB] (a32) at (6,-3) {};
			\node[superpeersgry, left color=IADB, IADB] (a33) at (4,-3) {};
			\node[superpeersgry, left color=IADB, IADB] (a41) at (9,-4) {};
			\node[superpeersgry, left color=IADB, IADB] (a42) at (7,-4) {};
			\node[superpeersgry, left color=IADB, IADB] (a43) at (5,-4) {};
			\node[superpeersgry, left color=IADB, IADB] (a44) at (3,-4) {};
			\node[superpeersgry, left color=IADB, IADB] (a51) at (10,-5) {};
			\node[superpeersgry, left color=IADB, IADB] (a52) at (8,-5) {};
			\node[superpeersgry, left color=IADB, IADB] (a53) at (6,-5) {};
			\node[superpeersgry, left color=IADB, IADB] (a54) at (4,-5) {};
			\node[superpeersgry, left color=IADB, IADB] (a55) at (2,-5) {};
			\node[superpeers, left color=IALB!20, IADB] (a61) at (11,-6) {};
			\node[superpeers, left color=IALB!20, IADB] (a62) at (9,-6) {};
			\node[superpeers, left color=IALB!20, IADB] (a63) at (7,-6) {};
			\node[superpeers, left color=IALB!20, IADB] (a64) at (5,-6) {};
			\node[superpeers, left color=IALB!20, IADB] (a65) at (3,-6) {};
			\node[superpeers, left color=IALB!20, IADB] (a66) at (1,-6) {};
			\node[void] (a71) at (12,-7) {};
			\node[void] (a72) at (10,-7) {};
			\node[void] (a73) at (8,-7) {};
			\node[void] (a74) at (6,-7) {};
			\node[void] (a75) at (4,-7) {};
			\node[void] (a76) at (2,-7) {};
			\node[void] (a77) at (0,-7) {};
			\node[void] (v1) at (0.3,-5) {};
			\node[void] (v3) at (0,-6) {};
			\path (a1) edge [bend right=20,->, IADB] (a21);
			\path (a1) edge [bend left=20,->, IADB] (a22);
			\path (a21) edge [bend right=20,->, IADB] (a31);
			\path (a21) edge [bend left=20,->, IADB] (a32);
			\path (a22) edge [bend right=20,->, IADB] (a32);
			\path (a22) edge [bend left=20,->, IADB] (a33);
			\path (a31) edge [bend right=20,->, IADB] (a41);
			\path (a31) edge [bend left=20,->, IADB] (a42);
			\path (a32) edge [bend right=20,->, IADB] (a42);
			\path (a32) edge [bend left=20,->, IADB] (a43);
			\path (a33) edge [bend right=20,->, IADB] (a43);
			\path (a33) edge [bend left=20,->, IADB] (a44);
			\path (a41) edge [bend right=20,->, IADB] (a51);
			\path (a41) edge [bend left=20,->, IADB] (a52);
			\path (a42) edge [bend right=20,->, IADB] (a52);
			\path (a42) edge [bend left=20,->, IADB] (a53);
			\path (a43) edge [bend right=20,->, IADB] (a53);
			\path (a43) edge [bend left=20,->, IADB] (a54);
			\path (a44) edge [bend right=20,->, IADB] (a54);
			\path (a44) edge [bend left=20,->, IADB] (a55);
			\path (a51) edge [bend right=20,->, IADB] (a61);
			\path (a51) edge [bend left=20,->, IADB] (a62);
			\path (a52) edge [bend right=20,->, IADB] (a62);
			\path (a52) edge [bend left=20,->, IADB] (a63);
			\path (a53) edge [bend right=20,->, IADB] (a63);
			\path (a53) edge [bend left=20,->, IADB] (a64);
			\path (a54) edge [bend right=20,->, IADB] (a64);
			\path (a54) edge [bend left=20,->, IADB] (a65);
			\path (a55) edge [bend right=20,->, IADB] (a65);
			\path (a55) edge [bend left=20,->, IADB] (a66);
			\path (a61) edge [bend right=20,->, IADB] (a71);
			\path (a61) edge [bend left=20,->, IADB] (a72);
			\path (a62) edge [bend right=20,->, IADB] (a72);
			\path (a62) edge [bend left=20,->, IADB] (a73);
			\path (a63) edge [bend right=20,->, IADB] (a73);
			\path (a63) edge [bend left=20,->, IADB] (a74);
			\path (a64) edge [bend right=20,->, IADB] (a74);
			\path (a64) edge [bend left=20,->, IADB] (a75);
			\path (a65) edge [bend right=20,->, IADB] (a75);
			\path (a65) edge [bend left=20,->, IADB] (a76);
			\path (a66) edge [bend right=20,->, IADB] (a76);
			\path (a66) edge [bend left=20,->, IADB] (a77);
			\path (a0) edge [->, very thick, IADB] (v1);
			\path (v2) edge [->, very thick, draw=IALB] (v3);
			\node[legend_general] at (-1.5,-3) {\small{bound = $2^k$}};
			\node[legend_general] at (-1.5,-6) {\small{bound = $2^{(k+1)}$}};
		\end{tikzpicture}
	}
	\caption{Bounded model-checking\cite{spijkerman2008marking}}
	\label{fig:screenshot006}
\end{figure}
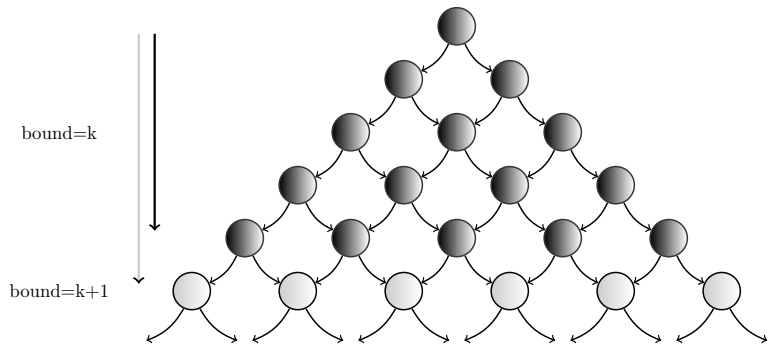

\paragraph{Bounded model-checking}
Bounded model-checking (BMC) has been proposed in \cite{biere1999symbolic,biere1999verifying} to deal with the complexity of model-checking and provide error traces. In this method, the length of the trace to be explored is limited via a fixed amount of states which will be indicated by $2^k \in {int}$, as illustrated in Figure \ref{fig:screenshot006}. Then, it checks through it to reveal error states. If an error location could not be reached inside the bound, the amount of $k$ will be increased and the process will be repeated until one error is found. The selected $k$ has to be large enough, otherwise, the method is not able to be completed \cite{biere2003bounded}. However, if $k$ is small enough, it outperforms BDD based model-checking \cite{clarke2001bounded}. Furthermore, discovering the $k$ involves minimal hands-on manipulations.  On the other hand BDD requires a great deal of hands-on effort to find an optimum ordering. In addition, BMC can handle much more clauses and variables than BDD methods \cite{biere2003bounded}.

 Bounded model-checking is the most important industrial application for the Boolean satisfiability (SAT) solver \cite{kautz1992planning}. SAT solver provides a platform for searching and reasoning based on propositional logic that is able to solve complex problems with millions of variables and constraints \cite{van2008handbook}.  SAT testing falls into NP-complete problems \cite{clarke2001bounded} and some recent works shows that using BMC in SAT leads to successfully verify security critical systems \cite{armando2014satmc}, concurrent systems \cite{phan2015concurrent}, multi agent systems \cite{wozna2016sat}, generating functional tests \cite{zhang2019software}, signal temporal logic \cite{ACM4}, and parallel and distributed systems \cite{ACM5}.

 BMC can also be based on satisfiability modulo theories (SMT) \cite{armando2009bounded}.  SMT allows to compress the formula when arrays and vectors are involved \cite{barrett2009satisfiability}. MBC was developed and used in a wide range of communities and domains \cite{cordeiro2012smt, alipour2016bounded, phan2015concurrent, inverso2020parallel}. An explicit version of BMC has been proposed in the literature in \cite{meulen2009breadth}.

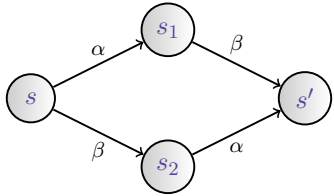
\begin{figure}[h!]
	\centering
	\resizebox{.7\columnwidth}{!}{
		\begin{tikzpicture}[auto, thick]
			\node[superpeers, left color=IALB!20, text=IADB, IADB] (d) at (0,0) {$s_2$};
			\node[superpeers, left color=IALB!20, text=IADB, IADB] (u) at (0,2) {$s_1$};
			\node[superpeers, left color=IALB!20, text=IADB, IADB] (l) at (-2,1) {$s$};
			\node[superpeers, left color=IALB!20, text=IADB, IADB] (r) at (2,1) {$s^\prime $};
			\draw[->,thick, IADB] (l) -- (u) node [pos=0.5,above,font=\footnotesize] {$\alpha$};
			\draw[->,thick, IADB] (l) -- (d) node [pos=0.5,below,font=\footnotesize] {$ \beta $};
			\draw[->,thick, IADB] (u) -- (r) node [pos=0.5,above,font=\footnotesize] {$ \beta $};
			\draw[->,thick, IADB] (d) -- (r) node [pos=0.5,below,font=\footnotesize] {$ \alpha $};
		\end{tikzpicture}
	}
	\caption{Some ordering results in a same state\cite{clarke1999state}}
	\label{fig:screenshot039}
\end{figure}

\paragraph{Partial order reduction}
This method attempts to cut down the state space of concurrent asynchronous systems \cite{godefroid1990using}. In asynchronous processes, interleaving models of executions must consider all possible orders of events for the sake of preventing the omission of important ones. Some of these ordering results in the same state,  as shown in Figure~\ref{fig:screenshot039}. To avoid this, partial order reduction avoids analyzing all sequences and considers only an incomplete set of events. Its methodology does not distinguish between traces that only differ by their orders. For example, in Figure~\ref{fig:screenshot039} to reach $s^\prime$ from $s$, it does not matter if $\alpha$ is run first or $\beta$. The set of events includes only representatives of enabled transition. Some approaches of partial order reduction is introduced in stubborn sets \cite{valmari1989stubborn}, ample sets \cite{alur1997partial}, persistent sets \cite{flanagan2005dynamic}, unfolding methods \cite{mcmillan1993approach}, and sleep sets \cite{clarke2008birth}.

Normally, the reduced model of partial order reduction is explicit and is produced by utilizing methods based on modified depth-first search \cite{valmari1990stubborn} or breadth-first search. It can be combined with other methods such as the on-the-fly model-checking \cite{peled1994combining} or symbolic model-checking \cite{alur1997partial}. This method reduces the memory usage and time requirements. One of the key factors  affecting the efficiency of these methods is the number of enabled transitions, which may change a predicate in the verified property \cite{robinson2001handbook}. In complex systems, the number of these transitions  increase, and fewer reductions can be constructed.

\paragraph{Abstraction}
Abstraction is a method to handle more state space by abstracting away the entire state space. It is based on the fact that states indicate computations in series of objects and obvious relationships that normally many similar behaviors are between them. The abstraction methods can interpret these objects in another universe of abstraction to avoid exploring all of them. However, the result of its execution must be the same as the original one \cite{cousot1996abstract}. In \cite{clarke1999model}, abstraction is defined as the following: let $S_i$ is $i_{th}$ state over the set of all states $S_1 \ldots S_n$. Abstraction will give a surjection $h = (h_1,\ldots,h_n)$ that groups and maps each state to corresponding abstract states.  In \cite{clarke1994model}, E. M. Clarke \textit{et al.} indicated that the abstraction can be done based on the following aspects: an equivalence modulo of an integer to address mathematical operations, symbolic abstraction, and single bit abstraction to address bit-by-bit logical operations.

As an example for the above kinds of abstraction, consider the following arithmetic modulo \cite{clarke1994model}:

\begin{center}\
	$(x\ mod\ i)+(y\ mod\ i)mod i\equiv x + y\ (mod\ i)$\\
	$(x\ mod\ i)-(y\ mod\ i)mod i\equiv x - y\ (mod\ i)$\\
	$(x\ mod\ i)(y\ mod\ i)\ mod\ i\equiv x\  y\ (mod\ i)$\\
\end{center}

To abstract the above modulo and determine the value of modulo $i$, we can use the values of modulo $i$ from the sub-expressions.

Another type of abstraction strategy based on storage reduction of states has been studied by G. J. Holzman \textit{et al.} \cite{holzmann2013coverage} that intends to minimize the size of the used memory during the construction of the state space. Additionally, another algorithm called \textit{cone of influence} is considered as an abstract method that removes all variables from the system model. The variables are idle or do not have any influence on the system properties \cite{clarke1999model}. Abstraction method has been successfully used to verify in many domains \cite{oortwijn2020abstraction} \cite{andre2019verification}.

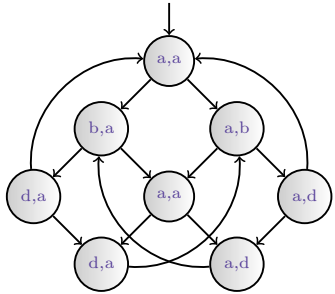
\begin{figure}[h!]
	\centering
	\resizebox{.7\columnwidth}{!}{
		\begin{tikzpicture}[auto, thick]
			\node[void] (v) at (3,1) {};
			\node[superpeers, left color=IALB!20, text=IADB, IADB] (a1) at (3,0) {\tiny{a,a}};
			\node[superpeers, left color=IALB!20, text=IADB, IADB] (a22) at (2,-1) {\tiny{b,a}};
			\node[superpeers, left color=IALB!20, text=IADB, IADB] (a21) at (4,-1) {\tiny{a,b}};
			\node[superpeers, left color=IALB!20, text=IADB, IADB] (a33) at (1,-2) {\tiny{d,a}};
			\node[superpeers, left color=IALB!20, text=IADB, IADB] (a32) at (3,-2) {\tiny{a,a}};
			\node[superpeers, left color=IALB!20, text=IADB, IADB] (a31) at (5,-2) {\tiny{a,d}};
			\node[superpeers, left color=IALB!20, text=IADB, IADB] (a41) at (4,-3) {\tiny{a,d}};
			\node[superpeers, left color=IALB!20, text=IADB, IADB] (a42) at (2,-3) {\tiny{d,a}};
			\path (v) edge [->, IADB] (a1);
			\path (a1) edge [->, IADB] (a21);
			\path (a1) edge [->, IADB] (a22);
			\path (a21) edge [->, IADB] (a31);
			\path (a21) edge [->, IADB] (a32);
			\path (a22) edge [->, IADB] (a32);
			\path (a22) edge [->, IADB] (a33);
			\path (a31) edge [->, IADB] (a41);
			\path (a32) edge [->, IADB] (a41);
			\path (a32) edge [->, IADB] (a42);
			\path (a33) edge [->, IADB] (a42);
			\path (a42) edge [bend right=50,->, IADB] (a21);
			\path (a41) edge [bend left=50,->, IADB] (a22);
			\path (a31) edge [bend right=50,->, IADB] (a1);
			\path (a33) edge [bend left=50,->, IADB] (a1);
		\end{tikzpicture}		
	}
	\caption{Mutual exclusion}
	\label{fig:screenshot001}
\end{figure}

\paragraph{Symmetry reduction methods}
Symmetry reduction method attempts to reduce the amount of state space. It is based on replacing sets of symmetrically similar states in a given model via a single representative class. Consider Figure~\ref{fig:screenshot001} of a mutual-exclusion for two components $a, b$ modeled by a $Kripke$ structure. There are many obvious symmetries between the components. For example, when component $a$ is in the critical section, component $b$ is waiting, equivalently, when the state in $b$ is in the critical section, $a$ is waiting. It is an adequate way of verification if we could find such equivalent states and check only one state from each class instead of checking all individual states.

A constructed model $M^\prime$ of system M under symmetry methods is called a quotient structure and a given property $fi$ holds for $M (fi \models M)$ if and only if, $fi$ holds for $M^\prime (fi \models M^\prime)$. Symmetry reduction methods have two difficulties: orbit problem, and constructive orbit problem. Orbit problem seeks to find if the two states $a$ and $\bar{a}$ are in the equal orbit \cite{clarke1998symmetry}. It is not known as NP-complete, but it is harder than isomorphism problems.

Constructive orbit problem (COP) is a representative function that replaces a set of symmetrically similar states in a given model by a single representative which is minimal. This problem falls into NP-hard problems \cite{babai1983canonical}.

\paragraph{Hash tables}
A set of effective reachable states in contrast to the amount of possibly reachable states of a given state space is few. All effective states are stored somewhere in the system's memory. One way to retrieve these sparse states, which can be visited before and will be needed several times during checking, is by using hash tables. A hash table is an alternative to direct addressing into an ordinary array. This property of  the hash table provides a simple way to examine an arbitrary state in a given array in O(1) time \cite{cormen2009introduction}. It can be considered as a yes-no method that is able to improve reachability analysis in verification processes.

 G. J. Holzman in \cite{holzmann1988improved} speeds up the process of generating an exhaustive list of all visited states during a search for errors and check new generations of states against all pre-analyzed states by a hash table. Through a hash table, the states can be accessed quickly and decrease the amount of state for checking. To compress information more, they ignored storing the hash key itself (states) and only used the hash value (the address computed) to identify a state. The hashing discipline has been used to improve state storage and state comparison. Another example is the SPIN model checker \cite{holzmann1998analysis} that utilizes hash functions.

\subsubsection{Bottom-up approach}
A verification process that can be done in a bottom-up manner prior to the entire state space construction. The state space will be checked bit by bit and the global properties can be deduced by combining their results. In the bottom-up approach, it deletes the states that are already checked from memory and can be re-verified when needed. On the other hand, the verification information from the verified-states can be reused without the need to re-verify.  In contrast to compositional verification, bottom-up approaches do not involve system decomposition. The following  two bottom-up approaches in alleviating SSE problem is discussed:

\paragraph{On-the-fly method} 
This method is an explicit method that is able to verify a system without storing the complete construction of the state space in memory. On-the-fly model-checking starts checking from an initial state and searches adjacent states to gain local knowledge about the state space in a stepwise manner. The key factor here is storing only the current path and  verification along with the construction of the system state space. In other words, it does not postpone the verification until the state space construction is completed. Therefore, the counterexample of the properties that do not hold can be found and generated as early as possible. This property is the most important advantage of the on-the-fly method \cite{schwoon2005note}.

Another advantage of the method is that it reduces the memory requirements substantially because it already eliminates visited states from memory. On the other hand, eliminating already verified states may increase the run time during error searching. Since this method does not store the already-visited states in memory and may need to regenerate them over and over, then the time of exploration grows dramatically.

 The methods themselves often employ depth-first-search (DFS) algorithms for searching through the state space. The run time of this method relies upon the number of states and  the number of transitions. The DFS algorithm is divided into two categories: Nested DFS and strongly connected component (SCC). Nested DFS, firstly searches for accepting states. Secondly, it searches for cycles around accepted states. Despite the memory efficiency of this algorithm, it may lead to finding a very long trace of a counterexample. In \cite{gastin2004minimization}, the authors proposed a method to achieve a minimal counterexample.

SCC-based on-the-fly methods find  a strongly connected component from an initial state to a given state. If any violation is found, then it produces a counterexample trace  that is strongly connected, and it includes at least one component. In comparison with nested DFS, it utilizes more memory,  a larger stack, and finally a longer counterexample. J. Geldenhuys introduced Tarjan's algorithm \cite{tarjan1972depth} to improve this kind of model-checking. Furthermore,  J. Geldenhuys \textit{et al}. \cite{geldenhuys2004tarjan} improved Tarjan's algorithm in terms of finding an accepting cycle sooner and producing a shorter counterexample. On-the-fly methods have been combined with other methods such as symmetry reduction \cite{patel2015fly}.

\paragraph{Incremental verification}
Incremental verification is one of those approaches that iteratively generates the SS of the system and verifies them until the overall properties of the system are satisfied. The key concept here is twofold: 1- preservation of the system properties when new increments   have been added; 2- providing an appropriate way to avoid re-verifying the system when new increments in the higher level of verification   have been added. Consequently, it reduces the  whole verification effort. Incremental verification is discussed in detail in section \ref{result}. \cite{bensalem2013incremental} is a research falling under this class and is discussed in section \ref{result}.

\subsubsection{Compositional Verification}
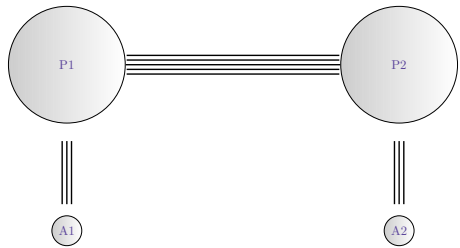
\begin{figure}[h!]
	\centering
	\resizebox{.7\columnwidth}{!}{
		\begin{tikzpicture}[auto, thick]
			\node[void] (vp1) at (4.8,0) {};
			\node[void] (vp2) at (11.2,0) {};
			\node[Giantpeers, left color=IALB!20, text=IADB, IADB] (p1) at (3,0) {\Huge P1};
			\node[Giantpeers, left color=IALB!20, text=IADB, IADB] (p2) at (13,0) {\Huge P2};
			\node[superpeers, left color=IALB!20, text=IADB, IADB] (a1) at (3,-5) {A1};
			\node[superpeers, left color=IALB!20, text=IADB, IADB] (a2) at (13,-5) {A2};
			\node[void] (vd1) at (8,-3) {};
			\node[void] (up1) at (3,-2.3) {};
			\node[void] (ua1) at (3,-4.2) {};
			\node[void] (up2) at (13,-2.3) {};
			\node[void] (ua2) at (13,-4.2) {};
			
			\path (p1) edge[very thick]   (p2);
			\path ([yshift=4pt]vp1.center) edge[very thick, IADB] ([yshift=4pt]vp2.center);
			\path ([yshift=-4pt]vp1.center) edge[very thick, IADB] ([yshift=-4pt]vp2.center);
			\path ([yshift=8pt]vp1.center) edge[very thick, IADB] ([yshift=8pt]vp2.center);
			\path ([yshift=-8pt]vp1.center) edge[very thick, IADB] ([yshift=-8pt]vp2.center);
			\path (up1.center) edge[very thick, IADB] (ua1.center);
			\path ([xshift=4pt]up1.center) edge[very thick, IADB] ([xshift=4pt]ua1.center);
			\path ([xshift=-4pt]up1.center) edge[very thick, IADB] ([xshift=-4pt]ua1.center);
			\path (up2.center) edge[very thick, IADB] (ua2.center);
			\path ([xshift=4pt]up2.center) edge[very thick, IADB] ([xshift=4pt]ua2.center);
			\path ([xshift=-4pt]up2.center) edge[very thick, IADB] ([xshift=-4pt]ua2.center);
		\end{tikzpicture}
	}
	\caption{ Abstraction}
	\label{fig:screenshot002}
\end{figure}

Compositional verification is a kind of divide-and-conquer approach which deals with SSE problem. It divides a large and complex problem into sub-problems and verifies each part separately. Contrary to its name, this method decomposes a given system into small components and then verifies the local properties of each component. One of the articles in this direction is \cite{phyo2019toward}. The verifying of local properties of the sub-systems contributes to the deduction of the entire system property. Obviously, by using this approach the whole SS does not need to be constructed and the sub-systems are not as big as the system itself. Consequently, the state-space volume will be significantly reduced. In addition, it provides more insight into the system interactions. Compositional verification has several alternative methods which are explained below:

\paragraph{Interface rule}
It makes an abstraction interface of component constraints and then  proves the preservation of each local  component by an interface rule. The idea behind using the interface abstraction is that in the composition process, only the properties are observable for other components should be checked. By hiding the rest of the properties, a huge amount of states will be reduced. Interface theory provides strong logical operations which are sound. The soundness of that is proved in \cite{berezin1998compositional}. Figure~\ref{fig:screenshot002} shows a general schema for the interface rule method. $P_1$ and $P_2$ are two processes or two components which is equipped with their interface rules $A_1$ and $A_2$.

\paragraph{Partitioned Transition Relations}
This method is based on the image of the states that produces a set of all successors of states $A$ and pre-image which produces the predecessor of the set of states  $A^\prime$ with a transition relation $T$ \cite{burch1991symbolic}. Let the sets $A$ and $T$ which are given by a Boolean formula, then the image of $A$ can be computed by the following formula:

\begin{equation}
	\exists m[A(m)  \wedge T(m,m')]
\end{equation}
which existential $m$ ($\exists m$) determines the quantification over all variables in the set of variables $m$.

In the first step, it constructs the transition relation $N_i$ of each component $i$ when exploring the system model and then composes all individual results to produce a global transition relation.  In this way, the global transition relation will never be constructed explicitly.  The formulas and steps for the synchronous systems  are as the following:

\begin{equation}
	\resizebox{0.85\columnwidth}{!}{
		$\underbrace{\exists m_{\rho(n-1)} [\cdots \underbrace{\exists m_{\rho(1)} [\underbrace{\exists m_{\rho(0)} [A(m) \wedge T_{\rho(0)}(m,m^\prime)]}_{A_1} \wedge T_{\rho(1)}(m,m^\prime)]}_{A_{2_{\vdots}}}\wedge \cdots \wedge T_{\rho (n-1)}(m,m^\prime)]}_{A_n}$
	}
\end{equation}

Clearly, each step depends on the previous step and the final partitioning strongly depends on the order that the variables are out qualified. It can be computed by optimum ordering of BDD (OBDD), however, finding an OBDD is complex and it needs special algorithms to find an optimum ordering. In \cite{burch1990sequential,berezin1998compositional}, the authors have presented an algorithm to compute an OBDD and improve the partition transition relations.

\paragraph{Lazy parallel composition}
For all processes in this method, a restricted transition relation will be created. The restricted transition is more concise than the global transition relation itself \cite{berezin1998compositional}. Let $R$ be a global transition relation and $S$ a set of states. $R^\prime$ can be a restricted transition relation for the image $s$ if it always satisfies the condition:

\begin{equation}
	R^\prime|_s = R|_s 
\end{equation}

The formula indicates that $R$ and $R^\prime$ concur on transitions starting from the state $s$, however,  $R^\prime$ has fewer nodes than $R$. This method simplifies the transition relation of each component by using the constraint operators before constructing the global transition relation.

\begin{equation}
	R^\prime = \bigwedge_{i .. n} constraint (R_i, S)
\end{equation}

$R^\prime$ must concur with the global relation transition $R$ in the set of states $S$. As a result, producing successors of $S$ by using the restricted transition $R^\prime$ produces the same result as using $R$. The total formula and steps that they take in comparison to the partial transition relation method is as the following:

\begin{equation}
	\exists m^\prime [A(m^\prime) \wedge \underbrace{(T_1(m,m^\prime ) |s}_{step 1} \wedge \underbrace{T_2(m,m^\prime ) |s}_{step 2}) ]
\end{equation}

In the formula, it is obvious that every step is independent, and it can be considered as an improvement of partition transition relations.

\paragraph{Assume-guarantee reasoning}
It has been proposed by Pnueli \cite{pnueli1985transition} where it verifies a single component of the system at a time. However, during the verification, a component needs to be associated with the assumption that the environment has a certain behavior, then if the other components of the system  guarantee the behavior, it can be deduced that the behavior holds true for the entire system. Thus, two kinds of properties should be checked: firstly, the specific assumptions about the environment behavior. Secondly, guarantees that the assumptions hold. The basic rule of assume-guarantee can be formulated as the following:

\begin{equation}
	\frac{\langle true\rangle M^\prime \langle g \rangle \; \; \; \; \;\langle M \langle f \rangle}{\langle true\rangle M \;\;\;\| \;\;\; M^\prime \langle f \rangle}
\end{equation}

This method is discussed in detail in section \ref{result}.

\begin{table*}[h]
	\centering	
	\setcounter{bottomnumber}{5}
	\resizebox{!}{!}{
		\begin{tabular}{c c c || c c c}
			\hline
			\hline
			\textbf{State Space Explosion Reduction} &  \textbf{explicit} & \textbf{implicit} & \textbf{State Space Explosion Reduction} &  \textbf{explicit} & \textbf{implicit}\\
			\hline
			Symbolic model-checking  &  & \checkmark & Genetic algorithm & \checkmark & \\
			Bounded model-checking &  & \checkmark & Random walk & \checkmark &  \\
			Garbage collection& \checkmark & & Ant colony &  \checkmark &  \\
			Partial order reduction & \checkmark & & Bloom filter & \checkmark & \\
			On-the-fly method & \checkmark && Incremental verification &\checkmark &\checkmark \\
			Expanding memory & \checkmark & \checkmark & Interface rules && \checkmark \\
			Symmetry reduction methods & \checkmark && Partition transition relation && \checkmark \\
			Abstraction & \checkmark && Lazy parallel composition && \checkmark \\
			Hash table & \checkmark && Assume-guarantee reasoning & \checkmark & \checkmark \\ 		
			\hline
		\end{tabular}			
	}
	\caption{Explicit and Implicit methods}
	\label{tab:implicit}
\end{table*}

\begin{table*}[h]
	\centering
	\begin{tabular}{c p{9cm}}
		\hline
		\hline
		\textbf{State Space Explosion Reduction Method} &     
		\textbf{Specific features} \\
		\hline
		Symbolic model-checking & \small{by choosing a sufficient encoding of the system model, symbolic model checking can support verification of $10^{20}$ states in practice \cite{burch1991symbolic}}. \\
		
		Bounded model-checking& it requires less by hand manipulation than other approaches like BDDs \cite{clarke2001bounded}.  \\
		
		Garbage collection & its memory management and fragmentation is on the fly that leads to reduce SSE problem; it avoids suspending pointers being created \cite{jones2016garbage}. \\
		
		Partial order reduction & it limits the searching of redundant interleaving that leads to search more states \cite{godefroid1990using}. \\
		
		On-the-fly method & it is able to verify individual traces rather than the whole state space; it is able to produce counterexamples  as early as possible \cite{schwoon2005note}.\\
		
		Expanding memory & external memory capacity is infinite and not expensive.  \\
		
		Symmetry reduction methods & it replaces a set of symmetrically similar states in a given model by a single representative that is minimal that the similar states and traces present only once \cite{clarke1998symmetry}.\\
		
		Abstraction & it handles more states by removing states with similar behaviors \cite{clarke1999model}.\\
		
		Compositional verification & the entire state space of the system will not be built completely, therefore by defining sufficient assumptions about the system and computational environment, SSE problem can reduce \cite{berezin1998compositional}.  \\
		
		Hash table & it does not need to store the whole states instead a hash table is used; it empowered nested depth-first-search to search and check more states \cite{holzmann1997state}\\
		
		Genetic algorithm & it uses an intelligent search instead of searching exhaustively into the entire state space \cite{yousefian2014heuristic}.  \\
		
		Random walk & its need for space is extremely low; it is able to be parallel \cite{pelanek2005enhancing, sivaraj2003random, 5368102}. \\
		
		Ant colony & coverage of all states is guarantee \cite{selvi2010comparative}. \\
		
		Incremental verification & it can be done before the system construction completes; the counterexamples can be found as early as possible [refer to section 4].  \\
		\hline
		\hline
	\end{tabular}
	\caption{Success factors of SSE reduction methods}
	\label{tab:2}
\end{table*}

\subsection{Comparison}
These methods for mitigating SSE problem so far have involved addressing the memory concern. For example, by using  a specific data structure like BDD, using heuristics, adding external memory, or dividing the problem into sub-problems. These methods are completely different and the only shared characteristics between them are the way the SS is represented. The SS representation can be divided into two main paradigms in model-checking, \textit{explicit and implicit}. Table~\ref{tab:implicit} compares the reviewed methods based on them.

The success factor and challenges of the reviewed methods are summarized in two separate tables. Table~\ref{tab:2} and \ref{tab:3} contains a list of the key features and challenges of each method that have been obtained in the literature and described in this section respectively.

\begin{table*}[]
	\small
	\centering
	\resizebox{!}{!}{
		\begin{tabular}{c p{9cm}}
			\hline
			\hline
			\textbf{State Space Explosion Reduction Methods} &   \textbf{Challenges} \\
			\hline
			Symbolic model-checking  & it uses BDDs which strongly depends on the ordering of its input variables \cite{bollig1996improving,bollig2014width}. \\

			Bounded model-checking& it is capable to find only trivial properties and unable to check systems contains deep loops \cite{d2008survey}; it can find long counterexample while short counterexamples are easier to understand \cite{schuppan2005shortest}. \\

			Garbage collection & it consuming computer resources \cite{zorn1993measured, hertz2005quantifying}; it is hard to predict the pauses that garbage collection has done \cite{220022}. \\

			Partial order reduction &  it does not sensitive in the ordering of traces \cite{clarke1999state}. \\

			On-the-fly method & it needs to regenerate already visited states which consequently lead to increase runtime \cite{jin2004compositional}. \\

			Expanding memory & it leads to slower memory access, therefore it needs a proper Input/Output algorithm  \cite{wu2015efficient}. \\

			Symmetry reduction methods &  it falls into NP-hard problems \cite{babai1983canonical}.  \\

			Abstraction & it needs a proper mapping function \cite{jin2004compositional}. \\

			Compositional verification &  breaking down a system is hard to do \cite{cobleigh2006breaking,cobleigh2008breaking}; verifying the systems with long chain circularity is difficult \cite{jin2004compositional}.  \\
			
			Hash table & it assigns a unique index number which may lead to a hash collision and missing the error states \cite{holzmann1997state}. \\
			
			Genetic algorithm, \\			
			Random walk, \\
			Ant colony &  these three methods do not guarantee the exploring all global states, therefore it does not guarantee to find global properties; time to coverage is uncertain \cite{selvi2010comparative}. \\			

			Bloom filter & there is no guarantee to find global properties, it may lead to false-positive results \cite{stern1995improved} .\\

			Incremental verification & it needs to generate rules to guarantee the preservation of properties and generating rules to avoid re-verifying the previous levels [refer to section \ref{result}]. \\
			\hline
			\hline
		\end{tabular}
	}
	\caption{Challenges of SSE reduction methods}
	\label{tab:3}
\end{table*}

\section{Results on handling SSE in CBSD}\label{result}
This section elaborates  on the results of handling SSE in CBSD that have been found in the research literature. The search is conducted on all  domains of CBSD as the focus of this research is to find the SSE reduction methods. The results indicate that despite the fact that deciding properties like liveness and deadlock-freeness in CBSD is NP-hard \cite{martens2006deciding, minnameier2006deadlock}; model-checking have been successfully utilized to evaluate this kind of properties of CBSD. A particular subset of CBSD verification is concerned with addressing SSE problem. The methods that have been explained in the previous section could be used in CBSD as well. However, regarding the selected studies in this research, the frequently used methods in CBSD  are assume-guarantee reasoning, interface rule, and incremental verification. The component-wise representation of the SS of these kinds of methods is one of the reasons  for their popularity in CBSD. Assume-guarantee and interface rule are subsets of compositional verification which falls into the divide and conquer category. The philosophy behind these two methods is dividing a system into some sub-components, addressing the local properties of a subset of components independently, and then deducing the entire properties of the system properties. Incremental verification falls into a bottom-up category that is able to exploit lower level verification information when small changes are applied, or components are added. Table~\ref{tab:4} represents the characteristics of these three methods. In this section, these methods are described followed by the key features and potential challenges.

\begin{table*}[!h]
	\centering
	\begin{tabular}{c p{9cm}}
		\hline
		\hline
		\textbf{State Space Explosion Reduction Methods} & \textbf{Challenges} \\
		\hline
		Assume-guarantee Reasoning & decomposing system, detecting appropriate assumption, developing new rules to determine the correctness of assumptions, circularity \\

		Interface-based verification & decomposing system, generating abstract constraint, refine the interface \\
		
		Incremental verification & preserving of system properties when new increments are added, avoiding of re-verifying in the high level of verification  \\
		\hline
		\hline
	\end{tabular}
	\caption{Challenges of SSE reduction methods in component-based verification}
	\label{tab:4}
\end{table*}

\textit{\textbf{Assume-guarantee reasoning-}} involves three steps starting with D (\textit{Three-D}): 1- \underline{D}ecomposing a given system $S$ into its sub components $C_1,C_2,\ldots,C_n$, 2- \underline{D}eriving assumption $A_i$ about the environment for each $C_i$, 3- \underline{D}efining rules to prove that properties of $C_i$ guarantees the requirements $\varphi$ of system $S$ under assumption $A_i$.

Step 1 breaks up the entire system into sub-components. Applying the divide and conquer methods over CBSD which is already composed of multiple components may facilitate the decomposition step, but it is still a tedious task. J. M Cobleigh et.al \cite{cobleigh2006breaking,cobleigh2008breaking} determined that finding an appropriate decomposition of a given system to verify by such methods is challenging.

Step 2 is the process of capturing the behavior that a given component $C_i$ collects about its environment $A_i$. The most important key point resulting in a successful assumption check is detecting the appropriate assumptions for every component. Thus, a challenging question arises here: How to detect an appropriate assumption? The assumptions have been traditionally generated by users that have hard limited the assume-guarantee reasoning practically. Some proposals have been proposed to develop the assumptions automatically such as using learning assumptions \cite{he2016learning, abd2018automated}.

Step 3 is defining the rules to prove that the sub-component $C_i$ guarantees its correct behavior under assumption $A_i$. Having rules to decide about the correctness of assumption is necessary. Thus, this question must be answered precisely: How to develop this kind of proof rules? The rules can be provided based on a set of theoretic operations \cite{chilton2012assume}.

Another challenge of this  method is shown  in Figure~\ref{fig:screenshot026}. Suppose a system with three components \textit{Sum}, \textit{Sub}, and \textit{Multiple}. It is compulsory to proof satisfaction of component \textit{Sum} to verify component \textit{Sub}. Likewise, verifying component \textit{Multiple} depends on \textit{Sum} and \textit{Sub}. This represents the problem of interdependent assumptions between components. Mutual interdependency between the components, like the dependency between component \textit{Sum} and \textit{Multiple} is another problem that is called \textit{circularity}. Tackling this problem needs a set of sound and complete rules. In \cite{jin2004compositional}, the researchers  prove that solving a long chain of circularity is difficult.

\tikzstyle{connector} = [->,thick]
\tikzstyle{connector2} = [thick]
\tikzstyle{Boxv}=[draw,black,
text=black!70]
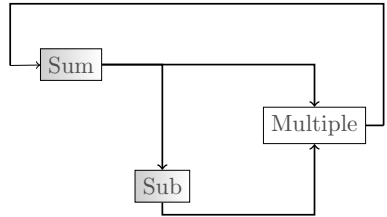
\begin{figure}[h!]
	\centering
	\resizebox{.7\columnwidth}{!}{
		\begin{tikzpicture}[path/.style={
				->,
				>=stealth,
				postaction=decorate
			}]
			\node[Box, left color=IALB!20, text=IADB, IADB, rounded corners] (sum) at (0,0) {Sum};
			\node[Box, left color=IALB!20, text=IADB, IADB, rounded corners] (sub) at (1.5,-2) {Sub};
			\node[Boxv, text=IADB, IADB, rounded corners] (mul) at (4,-1) {Multiple};
			\node[void] (v1) at (-2,0) {};
			\node[void] (v2) at (5,-1) {};
			\node[void] (v3) at (0.8,0.5) {};
			\node[void] (v4) at (3,-0.5) {\tiny{}};
			\node[void] (v5) at (3,-1.5) {\tiny{}};
			\node[void] (v10) at (-1,-0.15) {};
			\path (v10.north) edge [->, IADB] (sum);
			\draw [connector, IADB, rounded corners] (sub.south) -- ++(0,-.2cm) -| node [near start]
			{} (mul.south);
			\draw [connector, IADB, rounded corners] (sum.east) -| node [near start]
			{} (mul.north);
			\draw [connector, IADB, rounded corners] (sum.east) -| node [near end]{} (sub);
			\draw [connector2, IADB, rounded corners] (v2.east) -- ++(0,2cm) -| node [near end]
			{} (v10.north);
			\path (mul.east) edge [thick, IADB]  (v2.east);
		\end{tikzpicture}
	}
	\caption{Interdependency between components}
	\label{fig:screenshot026}
\end{figure}

Other works based on assume-guarantee reasoning involve an algorithm based on a prefix-closed set of traces \cite{chilton2012assume}, an algebraic theory \cite{muller2016component, chilton2013algebraic}, an algorithm named AGMC \cite{hoang2013assume}, an assume-guarantee verification for SOFA component model \cite{parizek2010assume}, interconnected systems \cite{al2020controller} and SSE reduction for time systems \cite{tran2018improvement, tran2019implementation, ghasemi2020compositional}.

\textit{\textbf{Interface rule reasoning-}} Interface rule which has been presented in \cite{clarke1999model} is a set of abstract constraints for each  single component in the systems. It restricts the behavior of components and assures the protection of local properties. The first key challenge is generating abstract constraints. This method can be used in compositional verification strategies by decomposing the interface of the system into sub-parts which  represent the global properties. Then, the composition of individual components should satisfy the global properties of the interface. The second challenge of this method is decomposition.

The interface rules must include appropriate information to fulfill the compositional verification goal. It can be either traditionally prepared by users manually or generated automatically. Some works based on this method are discussed below.

Y. Jin \cite{jin2004compositional} proposed a formal framework for specifying and verifying component-based systems based on \textit{interface automata (IA)}. IA has been used to describe interaction protocols of components and preserve the local properties of them to verify them independently. Another work on verifying component-based  systems via interface rule is presented by  A. Isazadeh \textit{et al.} \cite{isazadeh2009new}. They have proposed a formal model to specify the interface rule for communication protocols of components and then verify components according to this interface rule.

S. Quinton \textit{et al.} \cite{ben2010reasoning} have developed a framework for interface rule reasoning (in term of contract-based reasoning) for component-based design. In order to introduce such interface rule, they make benefit from a notation of I/A automata proposed by Henzinger and take into account sets of notations of constraints about composability and compatibility.

Another work presented in \cite{he2012component} on compositional verification of component-based in X-MAN. This work consists of two steps: 1- Vertical verification and Horizontal verification. Vertical verification guarantees that each atomic component satisfies its constraints (presented by an interface). 2- Horizontal verification which uses component constraint to verify the entire system. In this work, they suppose that the interface rules have been already attached to each component in the repository. Other works in interface based verification are \cite{sun2014verifying,cimatti2015contracts,howar2015trusting}.

\textit{\textbf{Incremental verification-}} as its name implies, incremental verification incrementally checks the system behaviour. It can be done during design process. If a CBSD is constructing incrementally, verification can be done during the construction and iteratively check properties of unfinished system until the system is completed. Thus, for incremental verification of CBSD, it is vital to support incremental construction. In \cite{lau2012incremental}, property feasibility of incremental construction have been proved. Let system $S$ be constructed by $inc_i$ increments from initial $inc_0$ where $inc_1 \subseteq inc_2 \subseteq inc_3 \ldots$ and $inc_i \subseteq inc_{i+1}$ which means $inc_{i+1}$ contains the behaviours of $inc_i$ \cite{lau2012incremental}. We say $\varphi$ is the properties of the system $S$ such that $\varphi_{inc_1} \subseteq  \varphi_{inc_2}  \subseteq \varphi_{inc_3} \subset \ldots$ where $\varphi_{inc_1} \subseteq  \varphi_{inc_2}$ means the properties that is satisfied in $inc_i$ is preserved in the next step of construction $inc_{i+1}$. Thus, preservation of the system properties when new increments are added is one of the key challenges in incremental verification.

Second, providing an appropriate way to avoid re-verifying the system when new increments in the higher level of verification are added is required. Verifying the whole system after every small change is not efficient. Therefore, providing a rule to deal with this issue can be considered as an effective improvement in formal verification, because it leads to reduce the whole verification effort significantly.

An incremental verification based on interaction invariants for component-based design has been proposed in \cite{bensalem2016component}. It is an invariant-based method for verification in the BIP component-based model. The interaction invariants involve locations of multiple components and express constraints on global SS induced via interactions. A method called binary behavioral constraints (BBC) has been proposed to symbolically compute the interaction invariants. The methods completely define the influence of interactions of a composite component on the other component's behavior To reuse those invariants when new increments  have been added they decompose the BBC and use two new methods to enhance scalability. Finally, the constraints and computations are represented by BDD. As discussed in Section 4, BDD strongly depends on input ordering and without it, the BDD may grow and quickly exceed the memory capacity. Despite proposing a sufficient technique to reduce the BDD, ordering BDD falls into NP-problems and does not have exact solutions \cite{bollig1996improving,bollig2014width}.

Other work concerning incremental verification for dynamic CBSD is defined in \cite{johnson2013incremental}. This work can verify CBSD whose components and structure change dynamically at runtime. The method, called INVEST, improved compositional verification by adding an incremental strategy to reverify a system after any removal, modification, and addition of components. Initially, the system will be verified by typical compositional verification and assume-guarantee reasoning. The incremental verification executes when any changes occur in the system.

\section{Discussion and Conclusion}\label{Section6}
This work on one hand reviews, briefly discusses, characterizes, and classifies existing methods of SSE reduction methods into five categories. On the other hand, it investigates the methods for alleviating SSE problem that have been utilized in CBSD. In section 3, RQ (1, 2, 3) have been answered. The state-of-the-art mitigation methods for SSE problem have been identified and explained. The key features and challenges of them are summarized as well. All these information have led to setting up a classification for common SSE reduction methods. RQs (4, 5) have been discussed in section 4. We demonstrated the common SSE reduction methods in CBSD and the potential challenges that have been obtained in the literature.

The general clue for this research is that despite proposing many methods for solving the bottleneck of model-checking, SSE problem still remains an obstacle in the worst case and has not been solved completely yet. This research provides a basis for many stakeholders such as component-based developers that need to select the most appropriate method for verifying their system, organizations that desire to create model checkers, and researchers seeking to set their research directions.

Having all the aforementioned information about SSE reduction methods, now we are in the position to discuss the proper method to be utilized in CBSD. Among the common methods for alleviating SSE problem in CBSD, is by using compositional verification in the form of either the assume-guarantee or interface rules. However, in such methods, the entire verification problem should be decomposed into the smaller task of its components and checked individually. Then, after decomposition, some difficulties such as interdependency between components, circularity, finding assumptions will arise. Utilizing such methods in order to verify CBSD obviously is limited by several issues.

On the other hand, applying incremental verification in CBSD may have some advantages. To begin with, such methods, omit the tedious task of breaking up the verification problem. Verifying a system in a bottom-up manner or bit-wise during incremental construction reduces the verification effort, rather than decomposing the system after the entire construction is finished and then verifying each part individually. The implementation of incremental construction and verification by \cite{bensalem2016component,lau2012incremental} has determined the possibility of this.

Another advantage of incremental verification is that counterexamples and error traces can be found as early as possible. In other words, the counterexamples can be created before the whole system is constructed. It is very useful to reveal error states before going through the higher levels of construction. However, all component-based models have yet to support incremental construction. Thus, providing a way to incrementally construct and verify component-based systems is a major direction for our future work. It might be possible by utilizing a component-based model with encapsulation mechanisms like what is presented in \cite{nejati2018putracom, nejatiputracom, 9122498, lau2005software}.

\section*{Acknowledgment}
We would like to thank the Ministry of Higher Education Malaysia and University Putra Malaysia for their generous grant FRGS/1/2015/ICT01/UPM/02/6.

\bibliographystyle{ieeetran}
\bibliography{ref}

\begin{IEEEbiography}[{\includegraphics[width=1in,height=1.25in,clip,keepaspectratio]{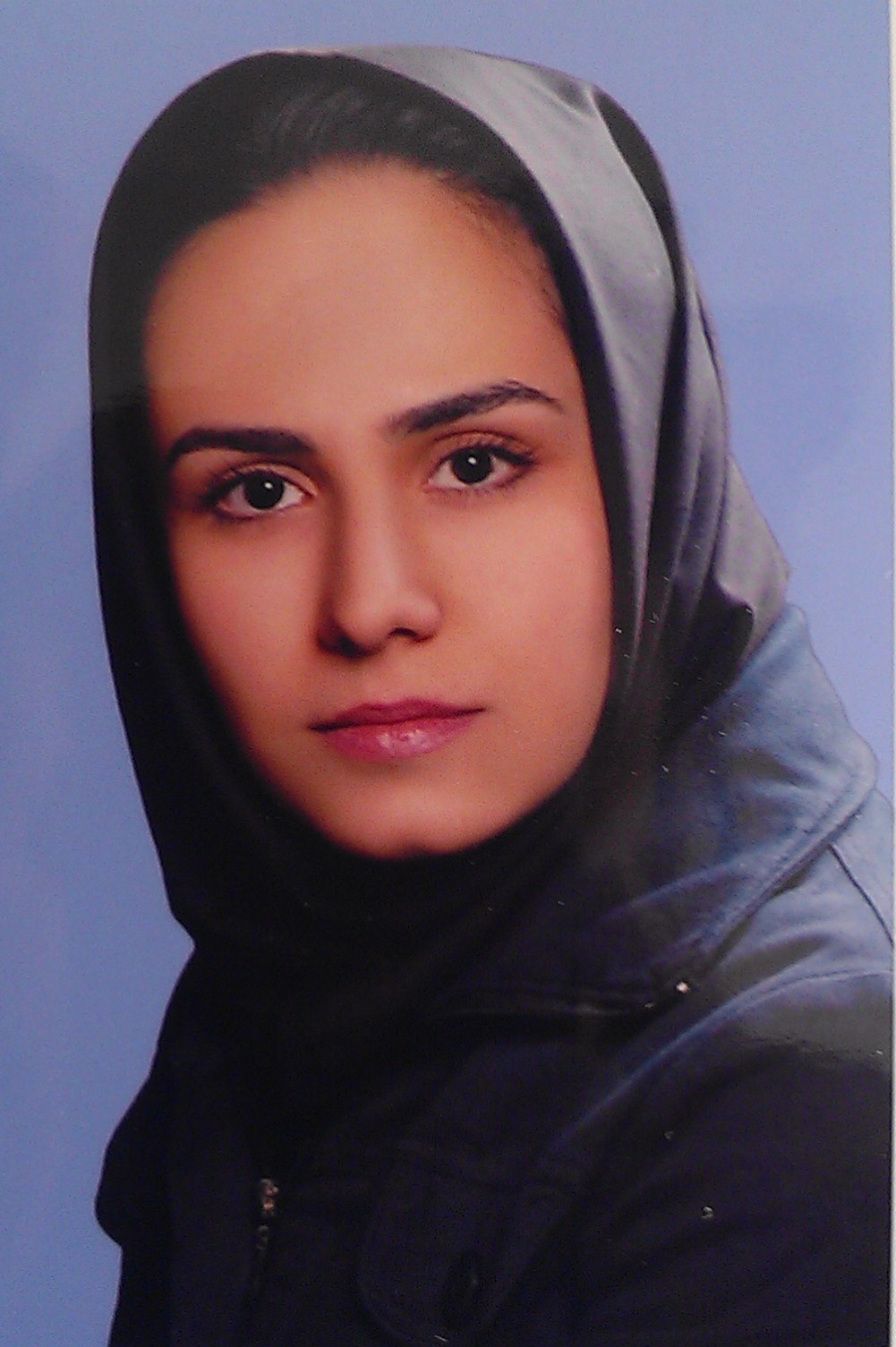}}]{} Faranak Nejati. She received the Ph.D. degree from the Department of Software Engineering, Faculty of Computer science and Information technology, University Putra Malaysia (UPM), Malaysia. She received the M.Sc in software Engineering from Tabriz University, Iran. Her research interests are Component-based software development, Software composition, Software connectors, Formal modeling, Formal verification, Model checking, Evolutionary algorithms, and Artificial Intelligence.	
\end{IEEEbiography}\vskip 0pt plus -1fil

\begin{IEEEbiography}[{\includegraphics[width=1in,height=1.25in,clip,keepaspectratio]{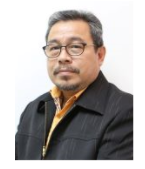}}]{}Abdul Azim Abd Ghani received the B.Sc in Mathematics/Computer Science from Indiana State University in 1984 and M.Sc in Computer Science from the University of Miami in 1985. He received the Ph.D. in Software Engineering from the University of Strathclyde in 1993. He is a Professor in Faculty of Computer Science and Information Technology, University Putra Malaysia. His research interests are software engineering, software measurement, software quality, and security in computing.
\end{IEEEbiography}\vskip 0pt plus -1fil

\begin{IEEEbiography}[{\includegraphics[width=1in,height=1.25in,clip,keepaspectratio]{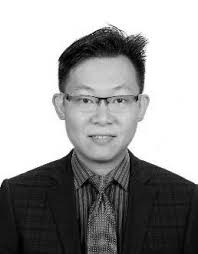}}]{}Ng Keng Yap, received B.Sc. in Comp. Sc. (UPM), M.Sc. (UPM), Ph.D (Manchester). He is a Senior Lecturer at University Putra Malaysia, Department of Software Engineering and Information System. His research interests are Software Architecture, Software Engineering (component-based, incremental software composition, software connectors, etc), Software Metric, Data Science and Business Analytics
\end{IEEEbiography} \vskip 0pt plus -1fil

\begin{IEEEbiography}[{\includegraphics[width=1in,height=1.25in,clip,keepaspectratio]{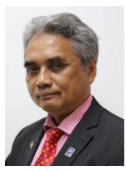}}]{} Azmi Bin Jafaar received B.Sc. (Mathematics and Computer Science) and M.Sc (Mathematics) from Indiana University, USA. His Ph.D is in Mathematical programming from Universiti Putra Malaysia in 1997. Currently, he is an associate professor at University Putra Malaysia, Department of Software Engineering and Information System. Also, he is a member of IAENGI international Association of Engineers and a member of Malaysian Mathematical Society (PERSAMA) since 1988. His research interests are Software Engineering, Discrete structures, mathematical programming, empirical methods in computer science.
	
\end{IEEEbiography}\vskip 0pt plus -1fil

\EOD
	
\end{document}